\documentclass[prd,preprint,nofootinbib]{revtex4}

\usepackage{amsmath}    
\usepackage{graphicx}   
\usepackage{slashed}
\usepackage{color}
\usepackage{feynmp}
\usepackage{psfrag}
\usepackage{xcolor}

\newcommand{\dd}{\mathrm{d}}

\newcommand{\als}{\alpha_s}
\newcommand{\ep}{\epsilon}

\newcommand{\nn}{\nonumber}

\def\fl#1{{#1}^\flat}

\def\spa#1.#2{\left\langle#1\,#2\right\rangle}
\def\spb#1.#2{\left[#1\,#2\right]}
\def\spaa#1.#2.#3{\langle\mskip-1mu{#1}
                  | #2 | {#3}\mskip-1mu\rangle}
\def\spbb#1.#2.#3{[\mskip-1mu{#1}
                  | #2 | {#3}\mskip-1mu]}
\def\spbab#1.#2.#3.#4{[\mskip-1mu{#1}
                  | #2 | #3 | {#4}\mskip-1mu]}
\def\spaba#1.#2.#3.#4{\langle\mskip-1mu{#1}
                  | #2 | #3 | {#4}\mskip-1mu\rangle}
\def\spab#1.#2.#3{\langle\mskip-1mu{#1}
                  | #2 | {#3}\mskip-1mu]}
\def\spba#1.#2.#3{[\mskip-1mu{#1}
                  | #2 | {#3}\mskip-1mu\rangle}

\begin{document}
\unitlength = 1mm
\title{One-loop Helicity Amplitudes for Top Quark Pair Production in Randall-Sundrum Model}
\author{Hua Xing Zhu}
\author{Chong Sheng Li}\email{csli@pku.edu.cn}
\author{Liang Dai}
\author{Jun Gao}
\author{Jian Wang}
\affiliation{Department of Physics and State Key Laboratory of Nuclear Physics
and Technology, Peking University, Beijing 100871, China}
\author{C.-P.Yuan}\email{yuan@pa.msu.edu}
\affiliation{Department of Physics and Astronomy, Michigan State University,
East Lansing, MI 48824, USA}
\begin{abstract}
In this paper, we show how to calculate analytically the one-loop
helicity amplitudes for the process $q\bar{q}\rightarrow t\bar{t}$
induced by KK gluon, using the spinor helicity formalism.  A minimal
set of Feynman rules which are uniquely fixed by gauge invariance
and the color representation of the KK gluon are derived and used in
the calculation. Our results can be applied to a variety of models
containing a massive color octet vector boson.
\end{abstract}
\maketitle

\section{Introduction}
\label{sec:intro}

Warped extra-dimension model proposed by Randall and
Sundrum~(RS)~\cite{Randall:1999ee} is a popular model that can solve
the hierarchy problem. When allowing the Standard Model~(SM) field
propagating in the extra dimension, the RS model provides many novel
points of view for some problems, like hierarchy of fermion masses
and the unification of gauge couplings.

The first signal of the RS model may be the observation of KK
gluon~(first Kaluza-Klein excitation mode of SM gluon) as a resonant
in the $t\bar{t}$ final state at the early LHC, simply because its
production rate is large compare to the other KK
particles~\cite{Lillie:2007yh}. Detailed study of total cross sections
and the invariant mass distribution of $t\bar{t}$ production induced
by KK gluon have been made in the literatures~\cite{Agashe:2006hk,Lillie:2007yh,Guchait:2007ux,Lillie:2007ve,Djouadi:2007eg,Frederix:2007gi,Baur:2008uv}
at the Leading Order~(LO) in QCD.

Recently CDF collaboration reported a large top quark forward-backward asymmetry with integrated luminosity $5.3\;\mathrm{fb}^{-1}$~\cite{Aaltonen:2011kc}. This $3.4\;\sigma$ discrepancy with SM predictions~\cite{Bowen:2005ap,Antunano:2007da} has motivated a lot of work in the theory community, {{cf. ref.}}~\cite{Krohn:2011tw} and references therein.

One attractive explanation of the large anomalous forward-backward
asymmetry is provided by the RS model. Within the RS framework, a realistic
model has been constructed that has been shown to be able to explain
the forward-backward asymmetry of top quark sector as well as bottom
quark sector simultaneously~\cite{Djouadi:2011aj}. This is realized
by a careful choice of fermion localizations, so that there is a
relatively light KK gluon with mass around $1.5$ TeV and large
parity violation in the first generation of quark couplings. The
authors of Ref.~\cite{Djouadi:2011aj} have shown at the LO in QCD
that the model they constructed can explain well the total asymmetry
observed at the Tevatron by CDF~\cite{Aaltonen:2011kc}, as well as
the asymmetry at large top quark pair invariant mass region and
large rapidity region~\cite{Aaltonen:2011kc}.

It's well known that LO prediction suffers from large scale
uncertainty, and therefore is not appropriate for {{precision}} measurement,
e.g., extraction of couplings between quark and KK gluon.
Furthermore, $t\bar{t}$ production is known to have large K-factor
at the Next-to-Leading
Order~(NLO) in QCD~\cite{Nason:1987xz,Beenakker:1988bq,Beenakker:1990maa}.
The large K-factor is partially taken into account in the
Ref.~\cite{Djouadi:2011aj} using a formalism proposed
in~\cite{Cao:2010zb}. Aiming at the precise prediction for the KK
gluon mediated $t\bar{t}$ production, Ref.~\cite{Bauer:2010iq}
calculated the interference of SM box diagrams and tree diagram of
$t\bar{t}$ production induced by KK gluon. Furthermore, Ref.~\cite{Allanach:2009vz} consider the effects of KK gluon on $t\bar{t}$ production by gluon-gluon fusion. It was found that the contributions, although non-zero, are highly suppressed. For this reason, we focus our attention the $q\bar{q}$ channel in the current work.

In this paper, we investigate how to calculate the complete NLO QCD
corrections to $t\bar{t}$ production induced by KK gluon, which has
not been reported in the previous literature. We first isolate a
minimal set of gauge invariant interactions and derive the relevant
Feynman rules for the KK gluon, the ghost of KK gluon and the 5th
component of the {{five-dimension~(5D)}} KK gluon field in $R_\xi$ gauge, and show in
detail how to renormalize the resulting one-loop amplitudes. Finally
we present the full one-loop helicity amplitudes for KK gluon induced
$t\bar{t}$ production, {{and the contribution from real emissions}} at the NLO is left
to be done in another work. As a by-product,
we also calculate the NLO total decay width of KK gluon in the large
$m_{KK}$ limit. A complete numerical result and phenomenology
discussion of the NLO corrections to $t\bar{t}$ production induced
by KK gluon will be presented elsewhere~\cite{hxzhu2011c}.

This paper is organized
as follows: in Sec.~\ref{sec:model} we briefly derive the relevant
Feynman rules used in our calculation. Details of the one-loop
calculation are presented in sec.~\ref{sec:helamp}. A discussion on
the result and a brief conclusion is given in
sec.~\ref{sec:conclusion}. All relevant Feynman rules can be found
in appendix~\ref{sec:appendix}.

\section{The Model}
\label{sec:model}
In this section we derive the Feynman rules relevant in our calculation. Part of the results in this section are well known in the literatures~\cite{Chang:1999nh,Huber:2000fh,Gherghetta:2000qt,Pomarol:1999ad,Davoudiasl:1999tf,Randall:2001gb,NovalesSanchez:2010yi,FloresTlalpa:2010rm}.
The RS construction is a slice of 5 dimensional Anti-de Sitter space. Since gravitational fluctuations play no role in the problem, we consider a fixed background metric of the form
\begin{equation}
  \label{eq:metric}
  ds^2 = \frac{1}{k^2 z^2} \left( \eta_{\mu\nu} dx^\mu dx^\nu - dz^2\right),
\end{equation}
where $z$ lives on the interval $[z_1=1/k,z_2=1/T]$. It's assumed that $k$ is of the order $M_{Pl}$, and $T$ is of the order TeV.

The action for a 5 dimensional massless gauge boson with $SU(3)$ gauge symmetry is
\begin{equation}
  \label{eq:action5d}
  S_{5D}=\int d^{4}xdz\sqrt{G}\left(-\frac{1}{2}\mathrm{Tr}F_{MN}F^{MN}\right),
\end{equation}
where the Roman indices $M,N$ run from $0\dots 3,5$. The field strength for the 5D gauge field is defined as
\begin{equation}
  \label{eq:fieldstrength}
  F_{MN}^{a}=\partial_{M}A_{N}^{a}-\partial_{N}A_{M}^{a}+g_{5}f^{abc}A_{M}^{b}A_{N}^{c},\qquad F_{MN}=F_{MN}^{a}T^{a},
\end{equation}
where $g_5$ is the gauge coupling constant in 5D. $T^a$ is the conventional Gell-mann matrix with normalization $\mathrm{Tr}[T^a T^b] = \frac{1}{2} \delta^{ab}$. Writing the action in terms of 4D components and the 5th component of the gauge field, we have
\begin{equation}
  \label{eq:explicitlag}
  S_{5D}=\int d^{4}x\int\frac{dz}{kz}\left[ -\frac{1}{4}F_{\mu\nu}^{a}F^{\mu\nu,a}-\frac{1}{2}F_{\mu5}^{a}F^{\mu5,a}\right].
\end{equation}
It can be seen from Eq.~(\ref{eq:explicitlag}) that there is mixing between the 4D components $A_\mu$ and $A_5$. To cancel the quadratic mixing terms, it's conventional to introduce the following bulk and boundary gauge fixing terms
\begin{gather}
  S_{GF,bulk}=\int d^{4}x\int\frac{dz}{kz}\left(-\frac{1}{2\xi}\right)\left[\partial^{\mu}A_{\mu}^{a}-\xi(kz)\partial_z\left(\frac{1}{kz}A_{5}^{a}\right)\right]^{2},
\\
S_{GF,boundary}=-\frac{1}{2\xi_{b}}\int d^{4}x\left[ \left.\left(\partial^{\mu}A_{\mu}^{a}+\xi_{b}\frac{1}{kz}A_{5}^{a}\right)^2\right|_{z=z_{2}}+\left.\left(\partial^{\mu}A_{\mu}^{a}-\xi_{b}\frac{1}{kz}A_{5}^{a}\right)^2\right|_{z=z_{1}}\right].
\end{gather}
A convenient gauge choice for the boundary terms is the unitary gauge $\xi_b \rightarrow \infty$, in which the boundary condition for the gauge field is
\begin{equation}
  \label{eq:boundaryc}
  \partial_z A^{\mu,a}|_{z=z_1,z_2}=0, \qquad
A^{a}_5 |_{z=z_1,z_2} = 0.
\end{equation}
As usual in quantizing spin-1 gauge fields, we also need to introduce ghost field to appropriately account for the degrees of freedom. Following the Faddeev-Popov procedure, the 5D ghost Lagrangian in $R_\xi$ gauge is
\begin{equation}
  \label{eq:ghostl}
  S_{5D,ghost}=\int d^{4}x\int\frac{dz}{kz}\bar{u}^{a}\left[-\partial^{\mu}\mathcal{D}_{\mu}+\xi\left(kz\right)\partial_{z}\frac{1}{kz}\partial_{z}\right]^{ab}u^{b},
\end{equation}
where $\mathcal{D}_\mu$ is the covariant derivative in adjoint representation:
\begin{equation}
  \label{eq:derivative}
  \left(\mathcal{D}_{M}\right)^{ab}=\delta^{ab}\partial_{M}+g_{5}f^{acb}A_{M}^{c}.
\end{equation}
The final action is then given by
\begin{equation}
  \label{eq:finall}
  S = S_{5D} + S_{GF,bulk} + S_{GF,boundary} + S_{5D,ghost}.
\end{equation}
To derive the relevant Feynman rules, we expand the gauge field in terms of a set of orthonormal KK {{modes}}:
\begin{eqnarray}
A_{\mu}\left(x,z\right)&=&\sqrt{k}\sum_{j=0}^{\infty}A_{\mu}^{(j)}(x)\chi_{j}(z),\nonumber\\
A_{5}\left(x,z\right)&=&\sqrt{k}\sum_{j=1}^{\infty}A_{5}^{(j)}(x)\frac{1}{m_{j}}\partial_{z}\chi_{j}(z),
\label{eq:kkdecomp}
\end{eqnarray}
where the orthonomal basis $\chi_j$ satisfies
\begin{equation}
\int\frac{dz}{z}\chi_{i}(z)\chi_{j}(z)=\delta_{ij}
\end{equation}
and  is determined by
\begin{eqnarray}
&&\partial_{z}\left(\frac{1}{z}\partial_{z}\chi_{j}(z)\right)+\frac{m_{j}^{2}}{z}\chi_{j}(z)=0 ,\nonumber\\
&&\partial_{z}\chi_{j}(z)|_{z=z_{1},z_{2}}=0.
\end{eqnarray}
The ghost field has similar KK decomposition in terms of 4D ghost field:
\begin{eqnarray}
u\left(x,z\right) &=&\sqrt{k}\sum_{j=0}^{\infty}u^{(j)}\left(x\right)\chi_{j}\left(z\right),\nonumber\\
\bar{u}\left(x,z\right) &=&\sqrt{k}\sum_{j=0}^{\infty}\bar{u}^{(j)}\left(x\right)\chi_{j}\left(z\right).
\end{eqnarray}
Substituting the expansion, Eq.~(\ref{eq:kkdecomp}), into the action, Eq.~(\ref{eq:explicitlag}), and integrating over the 5th dimension, we obtain the 4D Lagrangian that describes the interaction of various 4D field. The first few KK modes that are relevant to our discussion are $A^{(0)}_\mu$, $A^{(1)}_\mu$, $A^{(1)}_5$, $u^{(0)}$ and $u^{(1)}$. The propagator of these fields are given as{{,}} in $R_\xi$ gauge:
\begin{eqnarray}
\parbox{4cm}{
  \begin{fmffile}{prop1}
    \begin{fmfgraph*}(25,15)
      \fmfleft{i1}
      \fmfright{o1}
      \fmflabel{$a,\mu$}{i1}
      \fmflabel{$b,\nu$}{o1}
      \fmf{gluon,label=$p$,label.dist=0.5cm}{i1,o1}
    \end{fmfgraph*}
  \end{fmffile}
}
&=& -\frac{i\delta^{ab}}{p^2} \left(
 g^{\mu\nu} - \frac{p^\mu p^\nu}{p^2} (1-\xi) \right), \qquad A^{(0)}_\mu
\nn
\\
\parbox{4cm}{
  \begin{fmffile}{prop2}
    \begin{fmfgraph*}(25,15)
      \fmfleft{i1}
      \fmfright{o1}
      \fmflabel{$a,\mu$}{i1}
      \fmflabel{$b,\nu$}{o1}
      \fmf{dbl_curly,label=$p$,label.dist=0.5cm}{i1,o1}
    \end{fmfgraph*}
  \end{fmffile}
}
&=& -\frac{i\delta^{ab}}{p^2 - m^2_{KK}} \left(
 g^{\mu\nu} - \frac{p^\mu p^\nu}{p^2 - \xi m^2_{KK}} (1-\xi) \right),
\qquad A^{(1)}_\mu
\nn
\\
\parbox{4cm}{
  \begin{fmffile}{prop3}
    \begin{fmfgraph*}(25,15)
      \fmfleft{i1}
      \fmfright{o1}
      \fmflabel{$a$}{i1}
      \fmflabel{$b$}{o1}
      \fmf{dashes,label=$p$,label.dist=0.5cm}{i1,o1}
    \end{fmfgraph*}
  \end{fmffile}
}
&=& \frac{i\delta^{ab}}{p^2 - \xi m^2_{KK}},
\qquad A^{(1)}_5
\nn
\\
\parbox{4cm}{
  \begin{fmffile}{prop4}
    \begin{fmfgraph*}(25,15)
      \fmfleft{i1}
      \fmfright{o1}
      \fmflabel{$a$}{i1}
      \fmflabel{$b$}{o1}
      \fmf{dots_arrow,label=$p$,label.dist=0.5cm}{i1,o1}
    \end{fmfgraph*}
  \end{fmffile}
}
&=& \frac{i\delta^{ab}}{p^2},
\qquad u^{(0)}
\nn
\\
\parbox{4cm}{
  \begin{fmffile}{prop5}
    \begin{fmfgraph*}(25,15)
      \fmfleft{i1}
      \fmfright{o1}
      \fmflabel{$a$}{i1}
      \fmflabel{$b$}{o1}
      \fmf{dbl_dots_arrow,label=$p$,label.dist=0.5cm}{i1,o1}
    \end{fmfgraph*}
  \end{fmffile}
}
&=& \frac{i\delta^{ab}}{p^2 - \xi m^2_{KK}},
\qquad u^{(1)}
\end{eqnarray}
where we have identified the zero KK mode as the QCD gluon field, and denote the mass of first KK mode as $m_{KK}$. Without confusion, we also call the first KK mode as KK gluon occasionally.
It's a straightforward exercise to derive the Feynman vertices for these fields. For example, the vertex of 3 zero KK mode comes from the Lagrangian
\begin{equation}
  \label{eq:l3g}
  \mathcal{L}_{3g} =  \int \! \frac{\dd z}{kz}
\left( -\frac{1}{2} g_5 f^{abc} A^{(0),b}_\mu A^{(0),c}_\nu
(\partial_\mu A^{(0),a}_\nu - \partial_\nu A^{(0),a}_\mu ) \right) (\sqrt{k})^3 \chi^3_0(z).
\end{equation}
The $z$ integral can be trivially done since $\chi_0 \equiv \chi_0(z)$ is a constant{{, and}}
\begin{equation}
  \label{eq:g3int}
  \int\!\frac{\dd z}{kz} (\sqrt{k})^3 \chi^3_0(z) = \sqrt{k} \chi_0.
\end{equation}
It's immediately clear that one can identify the QCD gauge coupling as $g_s = \sqrt{k} \chi_0 g_5$, so that $\mathcal{L}_{3g}$ gives the conventional 3-point gluon interaction in QCD. One interesting feature of $\mathcal{L}_{3g}$ is that the resulting coupling is independent of profile of the KK mode in the 5th dimension. Actually one can derive a set of vertices that have this feature from the action in Eq.~(\ref{eq:finall}). Part of these vertices are just the conventional QCD vertices. The other part describes the interaction between the zero KK mode and the first KK mode, which can be found in the appendix.

Finally, we need to know the interaction between the KK mode and fermion, which is sensitive to the 5th dimension profile of {{the}} KK mode. This can be done by adding {{a}} term describing 5D fermion interaction with 5D gauge {{boson}} field into Eq.~(\ref{eq:finall}):
\begin{eqnarray}
S_{5D,fermion}&=&\int d^{5}x\sqrt{-g}\left\{ i\bar{\Psi}\Gamma^{M}D_{M}\Psi\right\} |_{\bar{\Psi}A\Psi\,\mathrm{piece}}\nonumber\\
&=&\int d^{4}x\int dz\left(\frac{1}{kz}\right)^{4}\bar{\Psi}\left[g_{5}\gamma^{\mu}A_{\mu}+ig_{5}\gamma^{5}A_{5}\right]\Psi\nonumber\\
&=&\int d^{4}x\int dz\left(\frac{1}{kz}\right)^{4}g_{5}\left\{ \psi\sigma^{\mu}A_{\mu}\bar{\psi}+\bar{\chi}\bar{\sigma}^{\mu}A_{\mu}\chi+i\left(-\psi\chi+\bar{\chi}\bar{\psi}\right)A_{5}\right\},
\label{eq:fermion}
\end{eqnarray}
where the covariant derivative is defined as
\begin{eqnarray}
D_{\mu}\Psi && =\left(\partial_{\mu}-\frac{i}{2z}\gamma_{\mu}\gamma_{5}\right)\Psi,\nonumber\\
D_{5}\Psi && =\partial_{z}\Psi,
\end{eqnarray}
and
\begin{equation}
  \Psi = \left(
    \begin{array}{c}
      \chi
\\
\bar{\psi}
    \end{array}
\right)
\end{equation}
is a Dirac spinor. The 5D fermion field can be expanded in terms of fermion KK {{modes}}:
\begin{eqnarray}
\chi\left(x,z\right) && =\sum_{j=0}g_{j}\left(z\right)\chi^{(j)}\left(x\right),\nonumber\\
\bar{\psi}\left(x,z\right) && =\sum_{j=1}f_{j}\left(z\right)\bar{\psi}^{(j)}\left(x\right),
\end{eqnarray}
where $g_j(z)$ and $f_j(z)$ are the fermion wave functions of the 5th dimension, with the normalization
\begin{equation}
\int dz\left(\frac{1}{kz}\right)^{4}g_{n}^{2}\left(z\right)=\int dz\left(\frac{1}{kz}\right)^{4}f_{n}^{2}\left(z\right)=1.
\end{equation}
Substituting the expansion into Eq.~(\ref{eq:fermion}), we derive the interaction between fermion zero mode and KK zero mode:
\begin{equation}
\int d^{4}xg_{s}\left\{ \psi^{(0)}\sigma^{\mu}A_{\mu}^{(0)}\bar{\psi}^{(0)}+\bar{\chi}^{(0)}\bar{\sigma}^{\mu}A_{\mu}^{(0)}\chi^{(0)}\right\}.
\end{equation}
This is just the conventional QCD interaction between fermion and gluon. For the interaction between the fermion zero mode and first KK mode, we have
\begin{equation}
\int d^{4}x\left\{ C_{R}\psi^{(0)}\sigma^{\mu}A_{\mu}^{(1)}\bar{\psi}^{(0)}+C_{L}\bar{\chi}^{(0)}\bar{\sigma}^{\mu}A_{\mu}^{(1)}\chi^{(0)}\right\},
\end{equation}
where the chiral couplings are defined as
\begin{equation}
C_{L}=\sqrt{k}g_{5}\int dz\left(\frac{1}{kz}\right)^{4}g_{0}^{2}\left(z\right)\chi_{1}\left(z\right),\qquad C_{R}=\sqrt{k}g_{5}\int dz\left(\frac{1}{kz}\right)^{4}f_{0}^{2}\left(z\right)\chi_{1}\left(z\right).
\end{equation}
There are also interaction terms between fermion zero mode and $A^{(1)}_5$:
\begin{eqnarray}
&&\int d^{4}x\int dz\left(\frac{1}{kz}\right)^{4}ig_{5}g_{0}\left(z\right)f_{0}\left(z\right)\left(-\psi^{(0)}\chi^{(0)}+\bar{\chi}^{(0)}\bar{\psi}^{(0)}\right)\sqrt{k}A_{5}^{(1)}\frac{1}{m_{KK}}\partial_{z}\chi_{1}\left(z\right)
\nonumber\\
=&&\int d^{4}x i g_{5}\frac{\sqrt{k}}{m_{KK}}\left(-\psi^{(0)}\chi^{(0)}+\bar{\chi}^{(0)}\bar{\psi}^{(0)}\right)A_{5}^{(1)}\int dz\left(\frac{1}{kz}\right)^{4}g_{0}\left(z\right)f_{0}\left(z\right)\partial_{z}\chi_{1}\left(z\right),
\end{eqnarray}
where all the $z$ dependencies have been written out explicitly. Also the 5th dimension wave function of gauge field $\chi_1(z)$ should not be confused with the chiral fermion field $\chi^{(0)}(x)$. Integrating by part over the $z$ integral, we obtain
\begin{eqnarray}
 && \int dz\left(\frac{1}{kz}\right)^{4}g_{0}\left(z\right)f_{0}\left(z\right)\partial_{z}\chi_{1}\left(z\right)\nonumber\\
= && \left[\left(\frac{1}{kz}\right)^{4}g_{0}f_{0}\partial_{z}\chi_{1}\right]_{z=z_{1}}^{z=z_{2}}-\int dz\left(\frac{1}{kz}\right)^{4}\left[-\frac{4}{z}g_{0}f_{0}+g_{0}^{\prime}f_{0}+g_{0}f_{0}^{\prime}\right]\chi_{1}\nonumber\\
= && \left[\left(\frac{1}{kz}\right)^{4}g_{0}f_{0}\partial_{z}\chi_{1}\right]_{z=z_{1}}^{z=z_{2}}-M_{0}\int dz\left(\frac{1}{kz}\right)^{4}\left[f_{0}^{2}-g_{0}^{2}\right]\chi_{1}\nonumber\\
= && \left[\left(\frac{1}{kz}\right)^{4}g_{0}f_{0}\partial_{z}\chi_{1}\right]_{z=z_{1}}^{z=z_{2}}-\frac{M_{0}}{g_5\sqrt{k}}\left(C_{R}-C_{L}\right),
\end{eqnarray}
where we have made use of the equation of motion of fermion field in the 5th dimension:
\begin{eqnarray}
f_{j}^{\prime}+M_{j}g_{j}-\frac{c+2}{z}f_{j} &&=0,\nonumber\\
g_{j}^{\prime}-M_{j}f_{j}+\frac{c-2}{z}g_{j} &&=0.
\end{eqnarray}
Here $M_j$ is the 4D mass of the $j$-th fermion KK mode, and $c$ is a bulk quark mass parameter, which doesn't appear in the interaction between fermion zero mode and $A^{(1)}_5$:
\begin{equation}
\int d^{4}x\left(-i\frac{M_{0}}{m_{KK}}\right)\left(C_{R}-C_{L}\right)\left(-\psi^{(0)}\chi^{(0)}+\bar{\chi}^{(0)}\bar{\psi}^{(0)}\right)A_{5}^{(1)}.
\end{equation}
The Feynman rules for the quark and first KK mode can be found in the appendix.

At this point, we have derived all the Feynman rules between the zero mode and the first KK mode that are uniquely determined by QCD gauge invariance, and the color representation of the KK mode. Vertices between the SM quark and first KK mode, though not fixed by gauge invariance, are also presented, since they are necessary for the process to happen. There exist other vertices which are not fixed by gauge invariance. For example, a vertex of 3 KK gluon can be derived from the Lagrangian, with a coupling sensitive to the 5th dimension profile of {{the}} KK mode. Such couplings might not be small; instead they are strong coupling in many cases. However we choose to omit these interactions in our calculation for several reasons:
\begin{itemize}
\item These couplings are usually strong, the meaning of perturbative expansion is not clear.
\item These couplings depend on the 5th dimension profile, thus are highly model dependent, and vary from model to model.
\item The effects of these couplings can be calculated separately, if desired.
\end{itemize}
 With the Feynman rules at hand, we are ready to explain {{the meaning of}} one-loop amplitudes for $q\bar{q} \rightarrow t \bar{t}$ in our calculation. These include the conventional SM QCD one-loop amplitudes, corrections of gluon self energy by loop of first KK mode, and the gluonic corrections to the LO process $q\bar{q} \rightarrow A^{(1)}_\mu \rightarrow t\bar{t}$. The amplitudes we consider have the features that they consist of a set of gauge invariant corrections, and are model independent~(in the sense that only the mass and color representation of the first KK mode matter). The structure of IR divergence of these amplitudes resemble  the SM QCD, and the IR divergences will be canceled when combining virtual corrections and real corrections. Thus all the low energy QCD effects are captured in our calculation, including the large threshold logarithms that usually dominate the NLO corrections~\cite{Ahrens:2011px}. The remaining diagrams that are not considered in this paper are both model dependent and IR finite. They can be calculated separately if needed. Similar consideration of calculating a subset of corrections can be found in {{ref.}}~\cite{Bohm:1982hr}.
\section{one-loop Helicity Amplitude}
\label{sec:helamp}
In this section we present the one-loop helicity amplitudes for $q\bar{q}\rightarrow t\bar{t}$, for both gluon induced and KK gluon induced processes. SM one-loop squared amplitudes for $t\bar{t}$ production are known for a long time~\cite{Nason:1987xz,Beenakker:1988bq,Beenakker:1990maa,Czakon:2008ii}. one-loop amplitudes with full helicity information are also known~\cite{Korner:2002hy,Badger:2011yu}. We have re-derived the SM one-loop amplitude for $q\bar{q}\rightarrow t \bar{t}$ and found complete agreement with those in {{ref.}}~\cite{Korner:2002hy}. Nevertheless we present them here for the sake of completeness.
\subsection{Convention}
Throughout our calculation, we adopt the Four-Dimensional Helicity (FDH) regularization scheme~\cite{Bern:2002zk}. Therefore the gauge coupling is defined in the FDH scheme. The conventional $\overline{\mathrm{MS}}$ scheme gauge coupling can be obtained by a finite renormalization~\cite{Kunszt:1993sd}
\begin{equation}
  \alpha^{\rm FDH}_s = \alpha^{\overline{\rm MS}}_s \left( 1 + \frac{\alpha^{\overline{\rm MS}}_s}{4\pi}\right).
\end{equation}
For simplicity, we do the calculation in 't Hooft-Feynman gauge, $\xi=1$. A common factor $C_\ep$ is omitted in all the result present below,
\begin{equation}
C_{\epsilon}=\frac{1}{\Gamma(1-\epsilon)}(4\pi)^{\epsilon}.
\end{equation}
Analytical continuation for the Mandelstam variables are defined as
\begin{gather}
  s \rightarrow s + i\varepsilon,
\nn
\\
 u \rightarrow u + i\varepsilon,
\nn
\\
t \rightarrow t + i \varepsilon.
\end{gather}
We use the modified spinor helicity method suitable for massive particles~\cite{Kleiss:1985yh} in our calculation. A recent application of this method can be found in {{ref.}}~\cite{Badger:2010mg}. As usual, massless spinor are denoted as
\begin{equation}
|i^{\pm}\rangle\equiv u_{\pm}\left(k_{i}\right)=v_{\mp}\left(k_{i}\right),\quad\langle i^{\pm}|\equiv\overline{u_{\pm}\left(k_{i}\right)}=\overline{v_{\mp}\left(k_{i}\right)}.
\end{equation}
Massive momenta {{are}} written as sum of two massless momenta:
\begin{equation}
p=p^{\ensuremath{\flat}}+\frac{M^{2}}{2p\cdot\eta}\eta,\quad p^{2}=M^{2},\quad\left(p^{\flat}\right)^{2}=\eta^{2}=0.
\end{equation}
Massive spinor can then be written as
\begin{eqnarray}
&&u_{\pm}\left(p,M;\eta,p^{\flat}\right)=\frac{\left(\slashed{p}+M\right)|\eta^{\mp}\rangle}{\langle p^{\flat\pm}|\eta^{\mp}\rangle},\quad\overline{u_{\pm}}\left(p,M;\eta,p^{\flat}\right)=\frac{\langle\eta^{\mp}|\left(\slashed{p}+M\right)}{\langle\eta^{\mp}|p^{\flat\pm}\rangle},\nonumber\\
&&v_{\pm}\left(p,M;\eta,p^{\flat}\right)=\frac{\left(\slashed{p}-M\right)|\eta^{\pm}\rangle}{\langle p^{\flat\mp}|\eta^{\pm}\rangle},\quad\overline{v_{\pm}}\left(p,M;\eta,p^{\flat}\right)=\frac{\langle\eta^{\pm}|\left(\slashed{p}-M\right)}{\langle\eta^{\pm}|p^{\flat\mp}\rangle},
\end{eqnarray}
where $\eta$ is an arbitrary reference light-like momenta. The arbitrariness of $\eta$ can be utilized to change the helicity of massive spinor:
\begin{equation}
\frac{\langle p^{\flat\mp}|\eta^{\pm}\rangle}{M}\overline{u_{\pm}}\left(p,M;p^{\flat},\eta\right)=\overline{u_{\mp}}\left(p,M;\eta,p^{\flat}\right),\quad\frac{\langle p^{\flat\mp}|\eta^{\pm}\rangle}{M}v_{\pm}\left(p,M;p^{\flat},\eta\right)=v_{\mp}\left(p,M;\eta,p^{\flat}\right).
\end{equation}
Therefore we only give results for amplitudes with a definite helicity configuration of massive quark, $\lambda_{3}=+,\,\lambda_{4}=+$, where $\lambda_3$ and $\lambda_4$ are the helicity of $t$ and $\bar{t}$, respectively.

For the process we consider in this paper, the amplitudes can be factorized into {{the}} product {{of}} a color factor and color stripped spinor products{{, \emph{i.e.},}} $\mathcal{M} = C \mathcal{A}$, where the color factor $C$ stands for some product of color matrixes. We will list below the color factor and spinor products respectively for each amplitude.
\subsection{SM Helicity Amplitude for $q\bar{q} \rightarrow t\bar{t}$}

\subsubsection{Results for LO Diagrams}
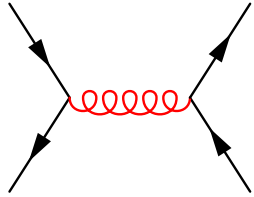
\begin{figure}[h!]
  \centering
    \begin{fmffile}{smtree}
    \begin{fmfgraph*}(40,25)
      \fmfleft{i2,i1}
      \fmfright{o2,o1}
      \fmf{quark}{i1,v1,i2}
      \fmf{quark}{o2,v2,o1}
      \fmf{gluon,fore=red}{v1,v2}
    \end{fmfgraph*}
  \end{fmffile}
  \caption{SM tree graph for $q\bar{q} \rightarrow t\bar{t}$.}
  \label{fig:smtree}
\end{figure}
The LO amplitudes~(Fig.~\ref{fig:smtree}) are straightforward to calculate. The color structure is
\begin{equation}
\left(T^{a}\right)_{i_{2}i_{1}}\left(T^{a}\right)_{i_{3}i_{4}}=\frac{1}{2}\delta_{i_{3}i_{1}}\delta_{i_{2}i_{4}}-\frac{1}{2}\frac{1}{N_{c}}\delta_{i_{2}i_{1}}\delta_{i_{3}i_{4}}.
\end{equation}
The Lorentz part is made of two structures of spinor products
\begin{eqnarray}
\mathcal{A}_{tree}\left(+,-,+,+\right) & = & \frac{8i\pi\alpha_{s}m_{t}}{s}\frac{\spa{\eta_4}.{1}\spab{\eta_3}.{\mathbf{3}}.{2}+\spa{\eta_3}.{1}\spab{\eta_4}.{\mathbf{4}}.{2}}{\spa{\fl 3}.{\eta_3}\spa{\eta_4}.{\fl 4}},
\nn
\\
\mathcal{A}_{tree}\left(-,+,+,+\right) & = & \frac{8i\pi\alpha_{s}m_{t}}{s}\frac{\spa{\eta_4}.{2}\spab{\eta_3}.{\mathbf{3}}.{1}+\spa{\eta_3}.{2}\spab{\eta_4}.{\mathbf{4}}.{1}}{\spa{\fl 3}.{\eta_3}\spa{\eta_4}.{\fl 4}},
\end{eqnarray}
where the boldface momenta {{denote}} massive {{particle momentum}} vector. At the LO, there is only vector current coupling $\bar{\psi}\gamma^{\mu}\psi$ at the massive quark vertex. At the NLO, however, magnetic-moment coupling $\bar{\psi}\left(i\sigma^{\mu\nu}q_{\nu}\right)\psi/\left(2m_{t}\right)$ is induced from loop diagram. Here we have defined $q=p_{3}+p_{4}$. For completeness we also list tree amplitudes for magnetic-moment interaction {{as follows:}}
\begin{eqnarray}
\mathcal{A}_{tree}^{(m)}\left(+,-,+,+\right) & = & -\frac{4i\pi\alpha_{s}}{sm_{t}}\frac{m_{t}^{2}\spb{2}.{1}\spa{\eta_3}.{1}\spa{\eta_4}.{1}+\spa{1}.{2}\spab{\eta_3}.{\mathbf{3}}.{2}\spab{\eta_4}.{\mathbf{4}}.{2}}{\spa{\fl 3}.{\eta_3}\spa{\eta_4}.{\fl 4}},
\nn
\\
\mathcal{A}_{tree}^{(m)}\left(-,+,+,+\right) & = & \frac{4i\pi\alpha_{s}}{sm_{t}}\frac{m_{t}^{2}\spb{2}.{1}\spa{\eta_3}.{2}\spa{\eta_4}.{2}+\spa{1}.{2}\spab{\eta_3}.{\mathbf{3}}.{1}\spab{\eta_4}.{\mathbf{4}}.{1}}{\spa{\fl 3}.{\eta_3}\spa{\eta_4}.{\fl 4}}.
\end{eqnarray}

\subsubsection{Results for Self-energy Diagrams}
\begin{figure}[h!]
    \begin{fmffile}{smsf1}
    \begin{fmfgraph*}(50,30)
      \fmfleft{i2,i1}
      \fmfright{o2,o1}
      \fmf{quark}{i1,v1,i2}
      \fmf{quark}{o2,v4,o1}
      \fmf{phantom,tension=0.3}{v1,v4}
      \fmffreeze
      \fmf{gluon,tension=5,fore=red}{v1,v2}
      \fmf{gluon,left,tension=1.5,fore=red}{v2,v3,v2}
      \fmf{gluon,tension=5,fore=red}{v3,v4}
    \end{fmfgraph*}
  \end{fmffile}
    \begin{fmffile}{smsf2}
    \begin{fmfgraph*}(50,30)
      \fmfleft{i2,i1}
      \fmfright{o2,o1}
      \fmf{quark}{i1,v1,i2}
      \fmf{quark}{o2,v3,o1}
      \fmf{gluon,tension=0.3,fore=red}{v1,v2,v3}
      \fmffreeze
      \fmf{gluon,fore=red}{v2,v2}
    \end{fmfgraph*}
  \end{fmffile}
    \begin{fmffile}{smsf3}
    \begin{fmfgraph*}(50,30)
      \fmfleft{i2,i1}
      \fmfright{o2,o1}
      \fmf{quark}{i1,v1,i2}
      \fmf{quark}{o2,v4,o1}
      \fmf{phantom,tension=0.3}{v1,v4}
      \fmffreeze
      \fmf{gluon,tension=5,fore=red}{v1,v2}
      \fmf{dots_arrow,left,tension=1.5}{v2,v3,v2}
      \fmf{gluon,tension=5,fore=red}{v3,v4}
    \end{fmfgraph*}
  \end{fmffile}
\\
\vspace{10mm}
    \begin{fmffile}{smsf4}
    \begin{fmfgraph*}(50,30)
      \fmfleft{i2,i1}
      \fmfright{o2,o1}
      \fmf{quark}{i1,v1,i2}
      \fmf{quark}{o2,v4,o1}
      \fmf{phantom,tension=0.3}{v1,v4}
      \fmffreeze
      \fmf{gluon,tension=5,fore=red}{v1,v2}
      \fmf{quark,left}{v2,v3}
      \fmf{quark,left,label=$q$}{v3,v2}
      \fmf{gluon,tension=5,fore=red}{v3,v4}
    \end{fmfgraph*}
  \end{fmffile}
    \begin{fmffile}{smsf5}
    \begin{fmfgraph*}(50,30)
      \fmfleft{i2,i1}
      \fmfright{o2,o1}
      \fmf{quark}{i1,v1,i2}
      \fmf{quark}{o2,v4,o1}
      \fmf{phantom,tension=0.3}{v1,v4}
      \fmffreeze
      \fmf{gluon,tension=5,fore=red}{v1,v2}
      \fmf{quark,left}{v2,v3}
      \fmf{quark,left,label=$t$}{v3,v2}
      \fmf{gluon,tension=5,fore=red}{v3,v4}
    \end{fmfgraph*}
  \end{fmffile}
  \caption{SM one-loop self energy graphs for $q\bar{q} \rightarrow t\bar{t}$.}
  \label{fig:smsf}
\end{figure}
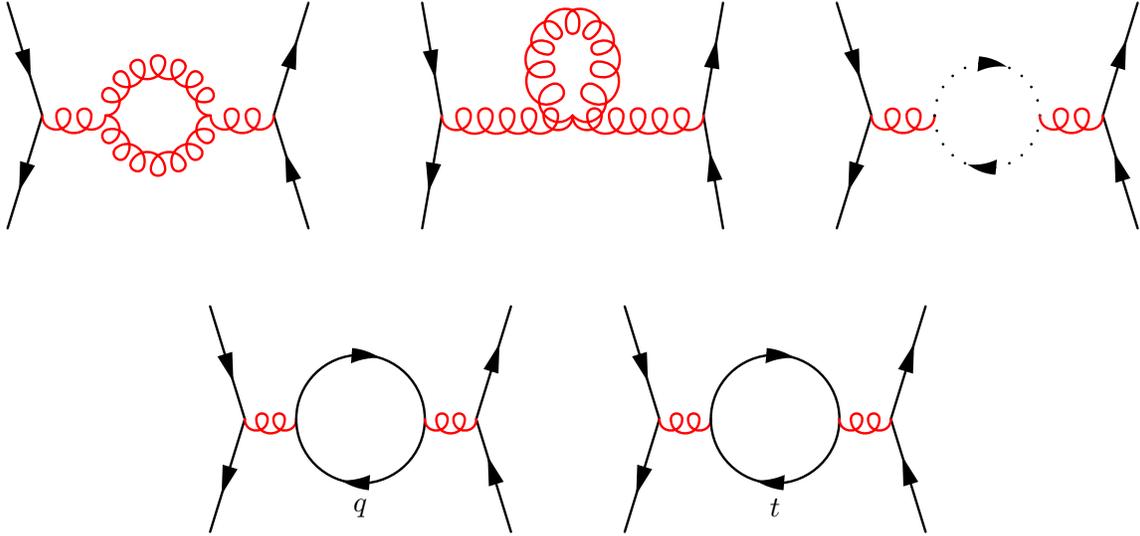
The amplitudes for self-energy diagrams~(Fig.~\ref{fig:smsf}) are proportional to the tree amplitudes. The color structure is identical to that of tree amplitudes. The Lorentz part is UV-divergent. The contributions from $n_{f}$ massless quark flavors, the massive top quark and the gluonic self-interactions are found to be
\begin{eqnarray}
\mathcal{A}_{sf}^{q}\left(\lambda_{1},\lambda_{2},\lambda_{3},\lambda_{4}\right) & = & \mathcal{A}_{tree}\left(\lambda_{1},\lambda_{2},\lambda_{3},\lambda_{4}\right)\frac{\alpha_{s}}{6\pi}n_{f}\left\{-\frac{1}{\epsilon_{\text{UV}}}+\ln\left(-\frac{s}{\mu^{2}}\right)-\frac{5}{3}\right\},
\nn
\\
\mathcal{A}_{sf}^{t}\left(\lambda_{1},\lambda_{2},\lambda_{3},\lambda_{4}\right) & = & \mathcal{A}_{tree}\left(\lambda_{1},\lambda_{2},\lambda_{3},\lambda_{4}\right)\frac{\alpha_{s}}{6\pi}\left\{-\frac{1}{\epsilon_{\text{UV}}}+\frac{2m_{t}^{2}}{s}\left(\beta\ln\left(\frac{\beta+1}{\beta-1}\right)-2\right)\right.
\nn
\\
&&\left.+\beta\ln\left(\frac{\beta+1}{\beta-1}\right)+\ln\left(\frac{m_{t}^{2}}{\mu^{2}}\right)-\frac{5}{3}\right\},
\nn
\\
\mathcal{A}_{sf}^{g}\left(\lambda_{1},\lambda_{2},\lambda_{3},\lambda_{4}\right) & = & \mathcal{A}_{tree}\left(\lambda_{1},\lambda_{2},\lambda_{3},\lambda_{4}\right)\frac{\alpha_{s}}{\pi}\left\{\frac{5}{4\epsilon_{\text{UV}}}-\frac{5}{4}\ln\left(-\frac{s}{\mu^{2}}\right)+\frac{7}{3}\right\}.
\end{eqnarray}
For $n_{f}=5$ massless quark flavors{{,}} the total self-energy diagram amplitude is simply
\begin{eqnarray}
\mathcal{A}_{sf}\left(\lambda_{1},\lambda_{2},\lambda_{3},\lambda_{4}\right) & = & \mathcal{A}_{tree}\left(\lambda_{1},\lambda_{2},\lambda_{3},\lambda_{4}\right)\frac{\alpha_{s}}{12\pi}\left\{\frac{3}{\epsilon_{\text{UV}}}-\frac{8m_{t}^{2}}{s}-5\ln\left(-\frac{s}{\mu^{2}}\right)+2\ln\left(\frac{m_{t}^{2}}{\mu^{2}}\right)\right.
\nn
\\
&&\left.+2\left(\frac{2m_{t}^{2}}{s}+1\right)\beta\ln\left(\frac{\beta+1}{\beta-1}\right)+8\right\}.
\end{eqnarray}
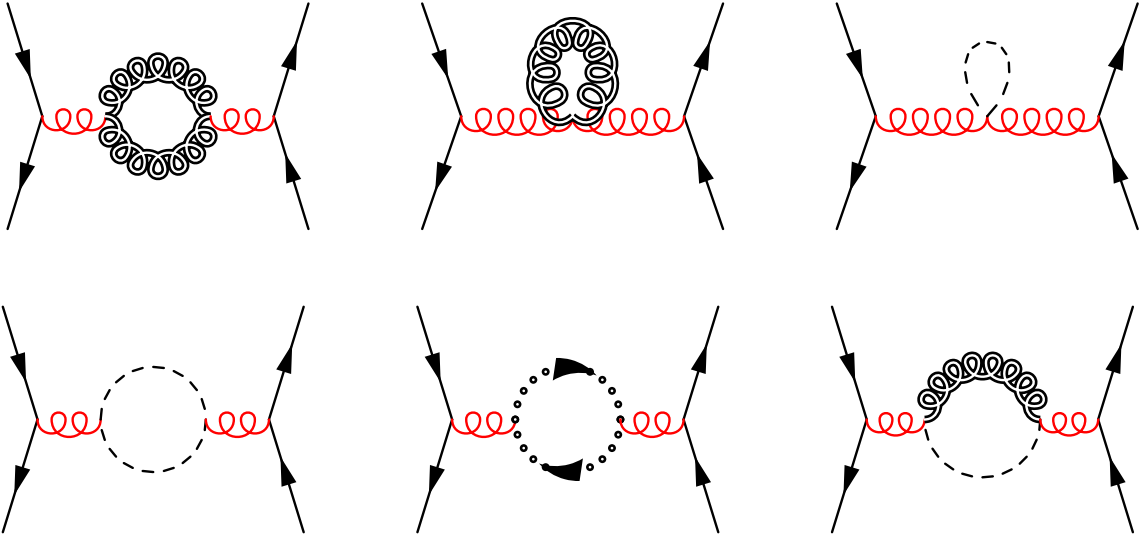
\begin{figure}[h!]
    \begin{fmffile}{smkksf1}
    \begin{fmfgraph*}(50,30)
      \fmfleft{i2,i1}
      \fmfright{o2,o1}
      \fmf{quark}{i1,v1,i2}
      \fmf{quark}{o2,v4,o1}
      \fmf{phantom,tension=0.3}{v1,v4}
      \fmffreeze
      \fmf{gluon,tension=5,fore=red}{v1,v2}
      \fmf{dbl_curly,left,tension=1.5}{v2,v3,v2}
      \fmf{gluon,tension=5,fore=red}{v3,v4}
    \end{fmfgraph*}
  \end{fmffile}
    \begin{fmffile}{smkksf2}
    \begin{fmfgraph*}(50,30)
      \fmfleft{i2,i1}
      \fmfright{o2,o1}
      \fmf{quark}{i1,v1,i2}
      \fmf{quark}{o2,v3,o1}
      \fmf{gluon,tension=0.7,fore=red}{v1,v2,v3}
      \fmffreeze
      \fmf{dbl_curly}{v2,v2}
    \end{fmfgraph*}
  \end{fmffile}
    \begin{fmffile}{smkksf3}
    \begin{fmfgraph*}(50,30)
      \fmfleft{i2,i1}
      \fmfright{o2,o1}
      \fmf{quark}{i1,v1,i2}
      \fmf{quark}{o2,v3,o1}
      \fmf{gluon,tension=0.7,fore=red}{v1,v2,v3}
      \fmffreeze
      \fmf{dashes}{v2,v2}
    \end{fmfgraph*}
  \end{fmffile}
\\
\vspace{10mm}
    \begin{fmffile}{smkksf4}
    \begin{fmfgraph*}(50,30)
      \fmfleft{i2,i1}
      \fmfright{o2,o1}
      \fmf{quark}{i1,v1,i2}
      \fmf{quark}{o2,v4,o1}
      \fmf{phantom,tension=0.3}{v1,v4}
      \fmffreeze
      \fmf{gluon,tension=5,fore=red}{v1,v2}
      \fmf{dashes,left,tension=1.5}{v2,v3,v2}
      \fmf{gluon,tension=5,fore=red}{v3,v4}
    \end{fmfgraph*}
  \end{fmffile}
    \begin{fmffile}{smkksf5}
    \begin{fmfgraph*}(50,30)
      \fmfleft{i2,i1}
      \fmfright{o2,o1}
      \fmf{quark}{i1,v1,i2}
      \fmf{quark}{o2,v4,o1}
      \fmf{phantom,tension=0.3}{v1,v4}
      \fmffreeze
      \fmf{gluon,tension=5,fore=red}{v1,v2}
      \fmf{dbl_dots_arrow,left,tension=1.5}{v2,v3,v2}
      \fmf{gluon,tension=5,fore=red}{v3,v4}
    \end{fmfgraph*}
  \end{fmffile}
    \begin{fmffile}{smkksf6}
    \begin{fmfgraph*}(50,30)
      \fmfleft{i2,i1}
      \fmfright{o2,o1}
      \fmf{quark}{i1,v1,i2}
      \fmf{quark}{o2,v4,o1}
      \fmf{phantom,tension=0.3}{v1,v4}
      \fmffreeze
      \fmf{gluon,tension=5,fore=red}{v1,v2}
      \fmf{dbl_curly,left,tension=1.5}{v2,v3}
      \fmf{dashes,left}{v3,v2}
      \fmf{gluon,tension=5,fore=red}{v3,v4}
    \end{fmfgraph*}
  \end{fmffile}
  \caption{KK gluon induced one-loop gluon self energy graph for $q\bar{q} \rightarrow t\bar{t}$.}
  \label{fig:smkksf}
\end{figure}
At one-loop level, the massive KK-gluon also enters the gluon propagator via gauge interactions with the gluon, and hence contributes to the gluon self-energy function. These diagrams~(Fig.~\ref{fig:smkksf}) give arise to an additional amplitude
\begin{eqnarray}
\mathcal{A}_{sf}^{KK}\left(\lambda_{1},\lambda_{2},\lambda_{3},\lambda_{4}\right) & = & \mathcal{A}_{tree}\left(\lambda_{1},\lambda_{2},\lambda_{3},\lambda_{4}\right)\frac{\alpha_{s}}{8\pi}\left\{\frac{9}{\epsilon _{\text{UV}}}-9 \ln \left(\frac{m_{KK}^2}{\mu ^2}\right)+8 \left(\frac{3 m_{KK}^2}{s}+2\right) \right.
\nn
\\
&&\left.-3 \left(\frac{4 m_{KK}^2}{s}+3\right)\tilde{\beta }\ln \left(\frac{\tilde{\beta }+1}{\tilde{\beta }-1}\right)\right\}.
\end{eqnarray}
Here we have defined $\tilde{\beta}=\sqrt{1-4m^2_{KK}/s}$.

The gluon wave-function renormalization constants enters the renormalization constants {{of}} the strong coupling. We renormalize the massless quark loops and gluonic loops in the $\overline{\mathrm{MS}}$ scheme, while for the massive top quark loop and KK gluon loop, on-shell scheme is adopted. Thus, massive particles are decoupled from the running of the strong coupling {{constant}}. Explicitly, various contributions to the gluon wave-function renormalization constants are
\begin{equation}
\delta Z_g=\delta Z_{g}^{(q),\overline{\mathrm{MS}}}+\delta Z_{g}^{(t),\mathrm{OS}}+\delta Z_{g}^{(g),\overline{\mathrm{MS}}}+\delta Z_{g}^{(KK),\mathrm{OS}},
\end{equation}
where
\begin{eqnarray}
\delta Z_{g}^{(q),\overline{\mathrm{MS}}} & = &\frac{\alpha_s}{\pi} n_f \left\{-\frac{1}{6 \epsilon _{\text{UV}}}\right\},
\nn
\\
\delta Z_{g}^{(t),\mathrm{OS}} & = &\frac{\alpha_s}{\pi}\left\{-\frac{1}{6 \epsilon _{\text{UV}}}+\frac{1}{6} \ln \left(\frac{m_t^2}{\mu ^2}\right)\right\},
\nn
\\
\delta Z_{g}^{(g),\overline{\mathrm{MS}}} & = &\frac{\alpha_s}{\pi}\left\{\frac{5}{4 \epsilon _{\text{UV}}}\right\},
\nn
\\
\delta Z_{g}^{(KK),\mathrm{OS}} & = &\frac{\alpha_s}{\pi}\left\{\frac{9}{8 \epsilon _{\text{UV}}}-\frac{9}{8} \ln \left(\frac{m_{KK}^2}{\mu ^2}\right)\right\}.
\end{eqnarray}
The corresponding counter-term diagram which renders the self-energy correction finite is
\begin{eqnarray}
\mathcal{A}_{sf}^{CT}\left(\lambda_{1},\lambda_{2},\lambda_{3},\lambda_{4}\right) & = & \mathcal{A}_{tree}\left(\lambda_{1},\lambda_{2},\lambda_{3},\lambda_{4}\right)\times\left(-\delta Z_g\right)
\nn
\\
& = & \mathcal{A}_{tree}\left(\lambda_{1},\lambda_{2},\lambda_{3},\lambda_{4}\right)\frac{\alpha_{s}}{\pi}\left\{-\frac{11}{8\epsilon _{\text{UV}}} -\frac{1}{6} \ln \left(\frac{m_t^2}{\mu ^2}\right) +\frac{9}{8} \ln \left(\frac{m_{KK}^2}{\mu ^2}\right)\right\}.
\nn
\\
\end{eqnarray}
\subsubsection{Results for Triangle Diagrams}
\begin{figure}[h!]
    \begin{fmffile}{smvt1}
    \begin{fmfgraph*}(50,30)
      \fmfleft{i2,i1}
      \fmfright{o2,o1}
      \fmf{quark}{i1,va,v1,vb,i2}
      \fmf{phantom}{o2,vc,v2,vd,o1}
      \fmf{gluon,fore=red}{v1,v2}
      \fmffreeze
      \fmf{quark}{o2,v2,o1}
      \fmf{gluon,fore=red}{va,vb}
    \end{fmfgraph*}
  \end{fmffile}
    \begin{fmffile}{smvt2}
    \begin{fmfgraph*}(50,30)
      \fmfleft{i2,i1}
      \fmfright{o2,o1}
      \fmf{phantom}{i1,va,v1,vb,i2}
      \fmf{phantom}{o2,vc,v2,vd,o1}
      \fmf{gluon,fore=red}{v1,v2}
      \fmffreeze
      \fmf{quark}{i1,va,vb,i2}
      \fmf{gluon,fore=red}{va,v1,vb}
      \fmf{quark}{o2,v2,o1}
    \end{fmfgraph*}
  \end{fmffile}
\\
\vspace{10mm}
    \begin{fmffile}{smvt3}
    \begin{fmfgraph*}(50,30)
      \fmfleft{i2,i1}
      \fmfright{o2,o1}
      \fmf{quark}{i1,va,v1,vb,i2}
      \fmf{phantom}{o2,vc,v2,vd,o1}
      \fmf{gluon,fore=red}{v1,v2}
      \fmffreeze
      \fmf{quark}{o2,v2,o1}
      \fmf{gluon,fore=red}{vc,vd}
    \end{fmfgraph*}
  \end{fmffile}
    \begin{fmffile}{smvt4}
    \begin{fmfgraph*}(50,30)
      \fmfleft{i2,i1}
      \fmfright{o2,o1}
      \fmf{phantom}{i1,va,v1,vb,i2}
      \fmf{phantom}{o2,vc,v2,vd,o1}
      \fmf{gluon,fore=red}{v1,v2}
      \fmffreeze
      \fmf{quark}{o2,vc,vd,o1}
      \fmf{gluon,fore=red}{vc,v2,vd}
      \fmf{quark}{i1,v1,i2}
    \end{fmfgraph*}
  \end{fmffile}
  \caption{SM one-loop triangle graphs for $q\bar{q} \rightarrow t\bar{t}$.}
  \label{fig:smvt}
\end{figure}
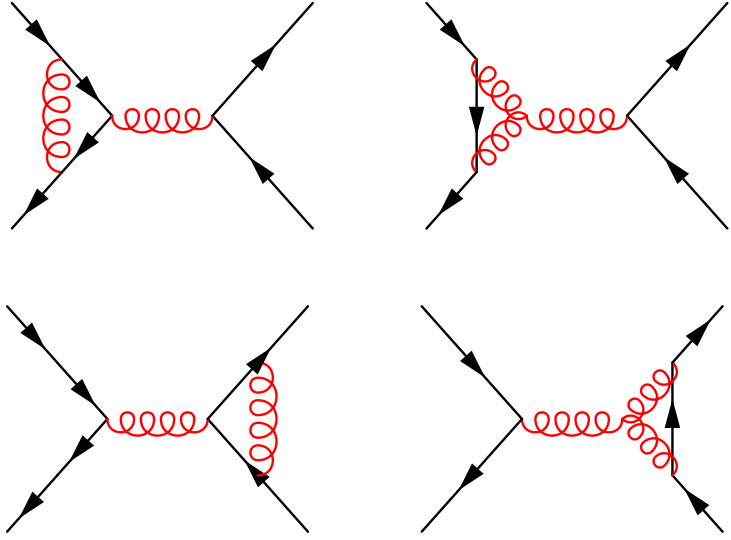
The color structure for the triangle diagram~(Fig.\ref{fig:smvt}) is the same as the tree amplitude. The Lorentz part of the amplitude can be divided into two parts. One is the vector current coupling induced part, which is UV- and IR- divergent at one-loop level. The other is the magnetic-moment type interaction induced part, which is free of divergence. Accordingly, we introduce two form factors $F_{1,2}\left(m_{t}^{2},s\right)$ and {{write}} the triangle diagram {{contribution as}}
\begin{equation}
\mathcal{A}_{vt}\left(\lambda_{1},\lambda_{2},\lambda_{3},\lambda_{4}\right)=\mathcal{A}_{tree}\left(\lambda_{1},\lambda_{2},\lambda_{3},\lambda_{4}\right)F_{1}\left(m_{t}^{2},s\right)+\mathcal{A}_{tree}^{(m)}\left(\lambda_{1},\lambda_{2},\lambda_{3},\lambda_{4}\right)F_{2}\left(m_{t}^{2},s\right).
\end{equation}
The one-loop triangle diagram of massless quark contributes only to the
vector current form factor $F_{1}\left(m_{t}^{2},s\right)${{:}}
\begin{eqnarray}
\mathcal{A}_{vt}^{q}\left(\lambda_{1},\lambda_{2},\lambda_{3},\lambda_{4}\right) & = & \mathcal{A}_{tree}\left(\lambda_{1},\lambda_{2},\lambda_{3},\lambda_{4}\right)\frac{\alpha_{s}}{4\pi}\left\{\frac{13}{3\epsilon_{\text{UV}}}+\frac{1}{3\epsilon_{\text{IR}}^{2}}-\frac{1}{\epsilon_{\text{IR}}}\left(\frac{1}{3}\ln\left(-\frac{s}{\mu^{2}}\right)+\frac{16}{3}\right)\right.
\nn
\\
&&\left.+\frac{1}{6}\ln^{2}\left(-\frac{s}{\mu^{2}}\right)+\ln\left(-\frac{s}{\mu^{2}}\right)-\frac{1}{3}\right\}.
\end{eqnarray}
The one-loop triangle diagram of massive quark contributes to both $F_{1}\left(m_{t}^{2},s\right)$ and $F_{2}\left(m_{t}^{2},s\right)${{:}}
\begin{eqnarray}
&&\mathcal{A}_{vt}^{t}\left(\lambda_{1},\lambda_{2},\lambda_{3},\lambda_{4}\right) = \mathcal{A}_{tree}\left(\lambda_{1},\lambda_{2},\lambda_{3},\lambda_{4}\right)\frac{\alpha_{s}}{4\pi}\left\{\frac{13}{3\epsilon_{\text{UV}}}+\frac{1}{\epsilon_{\text{IR}}}\frac{2m_{t}^{2}-s}{3s\beta}\ln\left(\frac{\beta+1}{\beta-1}\right)\right.
\nn
\\
&&\left.+\frac{1}{6 s \beta  s_1^2}\left[9 \beta  \ln \left(-\frac{s}{\mu ^2}\right) \left(s-16 m_t^2\right) s^2+36 \text{Li}_2\left(\frac{1}{1-\beta }\right) m_t^2 \left(8 m_t^2+s\right) s\right.\right.
\nn
\\
&&\left.\left.-72 \text{Li}_2(1-\beta ) m_t^2 \left(8 m_t^2+s\right) s+36 \text{Li}_2\left(\frac{1}{\beta +1}\right) m_t^2 \left(8 m_t^2+s\right) s \right.\right.
\nn
\\
&&\left.\left.-36 \text{Li}_2\left(\frac{\beta +1}{1-\beta }\right) m_t^2 \left(8 m_t^2+s\right) s-18 \text{Li}_2\left(\frac{s}{4 m_t^2}\right) m_t^2 \left(8 m_t^2+s\right) s\right.\right.
\nn
\\
&&\left.\left.+18 \text{Li}_2\left(\frac{4 m_t^2}{s}\right) m_t^2 \left(8 m_t^2+s\right) s-\beta  \ln \left(\frac{m_t^2}{\mu ^2}\right) \left(416 m_t^4-352 s m_t^2+35 s^2\right) s\right.\right.
\nn
\\
&&\left.\left.+2 \left(3 \pi ^2 \left(8 m_t^2+s\right) m_t^2+\beta  \left(352 m_t^4-212 s m_t^2+31 s^2\right)\right) s-2 \text{Li}_2\left(\frac{\beta -1}{2 \beta }\right) \left(2 m_t^2-s\right) s_1^2\right.\right.
\nn
\\
&&\left.\left.+2 \text{Li}_2\left(\frac{\beta +1}{2 \beta }\right) \left(2 m_t^2-s\right) s_1^2+\ln \left(\frac{\beta +1}{\beta -1}\right) \left(-s \left(3 s-8 m_t^2\right) s_1 \beta ^2\right.\right.\right.
\nn
\\
&&\left.\left.\left.-2 \ln (\beta ) \left(2 m_t^2-s\right) s_1^2-2 \ln \left(-\frac{s}{\mu ^2}\right) \left(2 m_t^2-s\right) s_1^2-\ln \left(-\frac{m_t^2}{s}\right) \left(2 m_t^2-s\right) s_1^2\right)\right]\right\}
\nn
\\
&&+ \mathcal{A}^{(m)}_{tree}\left(\lambda_{1},\lambda_{2},\lambda_{3},\lambda_{4}\right)\frac{\alpha_{s}}{6\pi}\frac{m^2_t}{\beta s^2_1}\left\{\ln \left(\frac{\beta +1}{\beta -1}\right) s_1 \beta ^2+9 \ln \left(-\frac{s}{\mu ^2}\right) \left(8 m_t^2+s\right) \beta \right.
\nn
\\
&&\left.-9 \ln \left(\frac{m_t^2}{\mu ^2}\right) \left(8 m_t^2+s\right) \beta -108 \text{Li}_2\left(\frac{1}{1-\beta }\right) m_t^2+216 \text{Li}_2(1-\beta ) m_t^2-108 \text{Li}_2\left(\frac{1}{\beta +1}\right) m_t^2 \right.
\nn
\\
&&\left.+108 \text{Li}_2\left(\frac{\beta +1}{1-\beta }\right) m_t^2+54 \text{Li}_2\left(\frac{s}{4 m_t^2}\right) m_t^2-54 \text{Li}_2\left(\frac{4 m_t^2}{s}\right) m_t^2-18 \left(\pi ^2 m_t^2+\beta  s_1\right)\right\}.
\end{eqnarray}
The renormalization constant {{of}} the strong coupling $g_s$ is given by
\begin{equation}
\delta Z_{g_s}=-\delta Z^{\overline{\mathrm{MS}}}_{\Gamma}-\delta Z^{\overline{\mathrm{MS}}}_q-\frac{1}{2}\delta Z^{\overline{\mathrm{MS}}}_g,
\end{equation}
where $\delta Z^{\overline{\mathrm{MS}}}_{\Gamma}$ is the UV-divergent part of the one-loop vertex function:
\begin{equation}
\delta Z^{\overline{MS}}_\Gamma = \frac{\als}{\pi}\frac{13}{12\ep_{\rm UV}},
\end{equation}
{{and}} $\delta Z^{\overline{\mathrm{MS}}}_q$ is just the UV-divergent part of the on-shell wave-function renormalization constant for massless quark. The on-shell wave-function renormalization constants for massless and massive quark are
\begin{eqnarray}
\delta Z^{\mathrm{OS}}_q & = & -\frac{\alpha_s}{3\pi}\left\{ \frac{1}{\epsilon_{\text{UV}}}-\frac{1}{\epsilon_{\text{IR}}} \right\},
\nn
\\
\delta Z^{\mathrm{OS}}_t & = & \frac{\alpha_s}{3\pi}\left\{ -\frac{1}{\epsilon_{\text{UV}}}-\frac{2}{\epsilon_{\text{IR}}}+3\ln \left(\frac{m_t^2}{\mu ^2}\right)-5\right\}.
\end{eqnarray}
The counter-term contributions that render both the massless quark vertex and the massive quark vertex {{to be}} UV-finite, respectively, {{are:}}
\begin{eqnarray}
\mathcal{A}_{vt}^{q,CT}\left(\lambda_{1},\lambda_{2},\lambda_{3},\lambda_{4}\right) & = & \mathcal{A}_{tree}\left(\lambda_{1},\lambda_{2},\lambda_{3},\lambda_{4}\right)\times\left(-\delta Z^{\overline{\mathrm{MS}}}_{\Gamma}-\delta Z^{\overline{\mathrm{MS}}}_q + \delta Z^{\mathrm{OS}}_q \right)
\nn
\\
& = & \mathcal{A}_{tree}\left(\lambda_{1},\lambda_{2},\lambda_{3},\lambda_{4}\right)\frac{\alpha_{s}}{\pi}\left\{-\frac{13}{12\epsilon_{\text{UV}}}+\frac{1}{3\epsilon_{\text{IR}}}\right\},
\nn
\\
\mathcal{A}_{vt}^{t,CT}\left(\lambda_{1},\lambda_{2},\lambda_{3},\lambda_{4}\right) & = & \mathcal{A}_{tree}\left(\lambda_{1},\lambda_{2},\lambda_{3},\lambda_{4}\right)\times\left(-\delta Z^{\overline{\mathrm{MS}}}_{\Gamma}-\delta Z^{\overline{\mathrm{MS}}}_t + \delta Z^{\mathrm{OS}}_t \right)
\nn
\\
& = & \mathcal{A}_{tree}\left(\lambda_{1},\lambda_{2},\lambda_{3},\lambda_{4}\right)\frac{\alpha_{s}}{\pi}\left\{-\frac{13}{12\epsilon_{\text{UV}}}-\frac{2}{3\epsilon_{\text{IR}}}+\ln \left(\frac{m_t^2}{\mu ^2}\right)-\frac{5}{3}\right\}.
\end{eqnarray}

\subsubsection{Results for Box Diagrams}
\begin{figure}[h!]
    \begin{fmffile}{smbox1}
    \begin{fmfgraph*}(50,30)
      \fmfleft{i2,i1}
      \fmfright{o2,o1}
      \fmf{quark}{i1,v1,v2,i2}
      \fmf{quark}{o2,v4,v3,o1}
      \fmf{gluon,tension=0.5,fore=red}{v1,v3}
      \fmf{gluon,tension=0.5,fore=red}{v2,v4}
    \end{fmfgraph*}
  \end{fmffile}
    \begin{fmffile}{smbox2}
    \begin{fmfgraph*}(50,30)
      \fmfleft{i2,i1}
      \fmfright{o2,o1}
      \fmf{quark}{i1,v1,v2,i2}
      \fmf{phantom}{o2,v4,v3,o1}
      \fmf{gluon,tension=0.5,fore=red}{v1,v3}
      \fmf{gluon,tension=0.5,fore=red}{v2,v4}
      \fmffreeze
      \fmf{quark}{o2,v3,v4,o1}
    \end{fmfgraph*}
  \end{fmffile}
  \caption{SM regular and cross box diagrams.}
  \label{fig:smbox}
\end{figure}
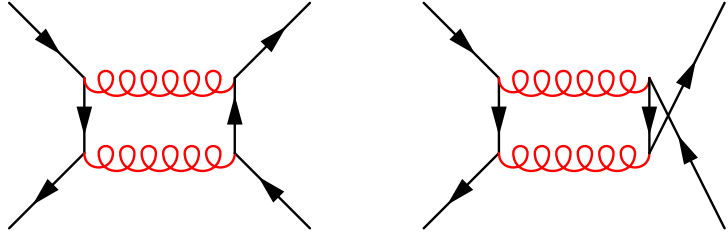
For each of the two helicity configurations for massless quarks, the
Lorentz parts of the box diagram~(Fig.\ref{fig:smbox}) amplitudes can be reduced to contain only
4 independent structure of spinor product. For the regular box diagram,
which is proportional to the color structure
\begin{equation}
\left(T^{a}T^{b}\right)_{i_{2}i_{1}}\left(T^{b}T^{a}\right)_{i_{3}i_{4}}=\frac{1}{4}\left(N_{c}-\frac{2}{N_{c}}\right)\delta_{i_{3}i_{1}}\delta_{i_{2}i_{4}}+\frac{1}{4}\frac{1}{N_{c}^{2}}\delta_{i_{2}i_{1}}\delta_{i_{3}i_{4}},
\end{equation}
we have found
\begin{eqnarray}
\mathcal{A}_{b1}\left(+,-,+,+\right)&=&\frac{4i\alpha_{s}^{2}m_{t}}{\spa{\fl 3}.{\eta_3}\spa{\eta_4}.{\fl 4}}\left\{ B_{1}\spa{\eta_4}.{1}\spab{\eta_3}.{\mathbf{3}}.{2}+B_{2}\spa{\eta_3}.{1}\spab{\eta_4}.{\mathbf{4}}.{2}\right.
\nn
\\
&&\left.+B_{3}\left[m_{t}^{2}\spb{2}.{1}\spa{\eta_{3}}.{1}\spa{\eta_{4}}.{1}+\spa{1}.{2}\spab{\eta_{3}}.{\mathbf{3}}.{2}\spab{\eta_{4}}.{\mathbf{4}}.{2}\right]\right\},
\nn
\\
\mathcal{A}_{b1}\left(-,+,+,+\right)&=&\frac{4i\alpha_{s}^{2}m_{t}}{\spa{\fl 3}.{\eta_3}\spa{\eta_4}.{\fl 4}}\left\{ B_{2}\spa{\eta_4}.{2}\spab{\eta_3}.{\mathbf{3}}.{1}+B_{1}\spa{\eta_3}.{2}\spab{\eta_4}.{\mathbf{4}}.{1}\right.
\nn
\\
&&\left.-B_{3}\left[m_{t}^{2}\spb{2}.{1}\spa{\eta_{3}}.{2}\spa{\eta_{4}}.{2}+\spa{1}.{2}\spab{\eta_{3}}.{\mathbf{3}}.{1}\spab{\eta_{4}}.{\mathbf{4}}.{1}\right]\right\},
\end{eqnarray}
where $B_{i},\, i=1,2,3${{,}} are auxiliary functions that depend on $m_{t}^{2}$
and the Mandelstam variables $s,\, t\,,u$. These functions can be
expressed in terms of the Passarino-Veltman one-loop functions
\begin{eqnarray}
B_{1} & = & 2D_{00}-t\left(D_{0}+D_{1}+D_{3}+D_{13}\right)+m_{t}^{2}(D_{0}+2D_{1}+D_{3}+D_{11}+D_{13}),
\nn
\\
B_{2} & = & 8D_{00}-t\left(D_{0}+D_{1}+D_{3}+2D_{13}\right)+m_{t}^{2}(D_{0}+2D_{1}+D_{3}+3D_{11}+2D_{13}),
\nn
\\
&&+2s\left(D_{2}+D_{12}+D_{22}+D_{23}\right),
\nn
\\
B_{3} & = & -D_{12},
\end{eqnarray}
where $D_{i},\, D_{ij}$ are defined as
\begin{eqnarray}
D_{i} & =\mathrm{PaVe}\left[i,\left\{ m_{t}^{2},m_{t}^{2},0,0,s,t\right\} ,\left\{ 0,m_{t}^{2},0,0\right\} \right],\\
D_{ij} & =\mathrm{PaVe}\left[i,j,\left\{ m_{t}^{2},m_{t}^{2},0,0,s,t\right\} ,\left\{ 0,m_{t}^{2},0,0\right\} \right].
\end{eqnarray}
We can define crossed auxiliary functions by exchange $t\longleftrightarrow u$, $\tilde{B}_{i}=-B_{i}\left(t\longleftrightarrow u\right)$. The amplitudes for the crossed box diagram, which is proportional to the color structure
\begin{equation}
\left(T^{a}T^{b}\right)_{i_{2}i_{1}}\left(T^{a}T^{b}\right)_{i_{3}i_{4}}=-\frac{1}{4}\frac{2}{N_{c}}\delta_{i_{3}i_{1}}\delta_{i_{2}i_{4}}+\frac{1}{4}\left(1+\frac{1}{N_{c}^{2}}\right)\delta_{i_{2}i_{1}}\delta_{i_{3}i_{4}}
\end{equation}
are
\begin{eqnarray}
\mathcal{A}_{b2}\left(+,-,+,+\right)&=&\frac{4i\alpha_{s}^{2}m_{t}}{\spa{\fl 3}.{\eta_3}\spa{\eta_4}.{\fl 4}}\left\{\tilde{B_{2}}\spa{\eta_4}.{1}\spab{\eta_3}.{\mathbf{3}}.{2}+\tilde{B_{1}}\spa{\eta_3}.{1}\spab{\eta_4}.{\mathbf{4}}.{2}\right.
\nn
\\
&&\left.+\tilde{B_{3}}\left[m_{t}^{2}\spb{2}.{1}\spa{\eta_{3}}.{1}\spa{\eta_{4}}.{1}+\spa{1}.{2}\spab{\eta_{3}}.{\mathbf{3}}.{2}\spab{\eta_{4}}.{\mathbf{4}}.{2}\right]\right\},
\nn
\\
\mathcal{A}_{b2}\left(-,+,+,+\right)&=&\frac{4i\alpha_{s}^{2}m_{t}}{\spa{\fl 3}.{\eta_3}\spa{\eta_4}.{\fl 4}}\left\{\tilde{B_{1}}\spa{\eta_4}.{2}\spab{\eta_3}.{\mathbf{3}}.{1}+\tilde{B_{2}}\spa{\eta_3}.{2}\spab{\eta_4}.{\mathbf{4}}.{1}\right.
\nn
\\
&&\left.-\tilde{B_{3}}\left[m_{t}^{2}\spb{2}.{1}\spa{\eta_{3}}.{2}\spa{\eta_{4}}.{2}+\spa{1}.{2}\spab{\eta_{3}}.{\mathbf{3}}.{1}\spab{\eta_{4}}.{\mathbf{4}}.{1}\right]\right\}.
\end{eqnarray}
Next we give explicit expressions for $B_{i}$, suitable for general complex arguments. First we introduce some notations
\begin{equation}
s_{1}=s-4m_{t}^{2},\quad t_{1}=m_{t}^{2}-t,\quad u_{1}=m_{t}^{2}-u,\quad K=m_{t}^{4}-tu,\quad\beta=\sqrt{1-4m_{t}^{2}/s}.
\end{equation}
$B_{3}$ is finite
\begin{eqnarray}
B_{3} & = & -\frac{1}{K^{2}s_{1}s\beta}\left\{\text{Li}_{2}\left(\frac{1}{1-\beta}\right)-2\text{Li}_{2}(1-\beta)+\text{Li}_{2}\left(\frac{1}{\text{\ensuremath{\beta}}+1}\right)-\text{Li}_{2}\left(\frac{\beta+1}{1-\beta}\right)+\frac{1}{2}\text{Li}_{2}\left(\frac{4m_{t}^{2}}{s}\right)\right.
\nn
\\
&&\left.-\frac{1}{2}\text{Li}_{2}\left(\frac{s}{4m_{t}^{2}}\right)+\frac{\pi^{2}}{6}\right\}C_{1} -\frac{1}{12K^{2}s_{1}}\left\{(-3s_{1}\ln^{2}\left(-\frac{s}{\mu^{2}}\right)t_{1}^{2}-6s_{1}\ln^{2}\left(\frac{t_{1}}{\text{\ensuremath{\mu^{2}}}}\right)t_{1}^{2}\right.
\nn
\\
&&+3s_{1}\ln^{2}\left(\frac{m_{t}^{2}}{\mu^{2}}\right)t_{1}^{2}-4\pi^{2}s_{1}t_{1}^{2}+12s_{1}\text{Li}_{2}\left(-\frac{t}{t_{1}}\right)t_{1}^{2}+12Ks\ln\left(\frac{t_{1}}{\text{\ensuremath{\mu^{2}}}}\right)
\nn
\\
&&+\ln\left(-\frac{s}{\text{\ensuremath{\mu^{2}}}}\right)\left(12s_{1}\ln\left(\frac{t_{1}}{\mu^{2}}\right)t_{1}^{2}+12K\left(m_{t}^{2}+t\right)\right)
\nn
\\
&&\left.+\ln\left(\frac{m_{t}^{2}}{\text{\ensuremath{\mu^{2}}}}\right)\left(12K\left(m_{t}^{2}+u\right)-6s_{1}t_{1}^{2}\ln\left(-\frac{s}{\text{\ensuremath{\mu^{2}}}}\right)\right)\right\},
\end{eqnarray}
where the coefficient $C_{1}$ is a polynomial of $m_{t}^{2},\, s,\, t,\, u$
\begin{equation}
C_{1}=4m_{t}^{8}-6sm_{t}^{6}-12tm_{t}^{6}+s^{2}m_{t}^{4}+12t^{2}m_{t}^{4}+8stm_{t}^{4}-4t^{3}m_{t}^{2}-2st^{2}m_{t}^{2}-2s^{2}tm_{t}^{2}-s^{2}t^{2}.
\end{equation}
The other two functions $B_{1},B_{2}$ have the same IR-divergent part so that divergences are proportional to the tree amplitudes
\begin{equation}
B_{1,2}=B_{1,2}^{fin}+\frac{1}{s}\left\{-\frac{1}{\epsilon_{\text{IR}}^{2}}+\frac{1}{\epsilon_{\text{IR}}}\left(2\ln\left(\frac{t_{1}}{\mu^{2}}\right)-\ln\left(\frac{m_{t}^{2}}{\mu^{2}}\right)\right)\right\},
\end{equation}
where $B_{1,2}^{fin}$ are the finite parts, respectively. We also give explicit expressions for both finite {{parts}}. For $B_{2}^{fin}${{,}}
we define coefficients
\begin{eqnarray}
C_{2} & = & 4m_{t}^{8}-7tm_{t}^{6}-3um_{t}^{6}+5t^{2}m_{t}^{4}+u^{2}m_{t}^{4}+4tum_{t}^{4}-t^{3}m_{t}^{2}-2tu^{2}m_{t}^{2}-3t^{2}um_{t}^{2}+2t^{2}u^{2},
\nn
\\
C_{3} & = & m_{t}^{4}-3tm_{t}^{2}-um_{t}^{2}+t^{2}+2tu,
\nn
\\
C_{4} & = & 2tm_{t}^{8}-6um_{t}^{8}-7t^{2}m_{t}^{6}+u^{2}m_{t}^{6}+6tum_{t}^{6}-t^{3}m_{t}^{4}+u^{3}m_{t}^{4}+5tu^{2}m_{t}^{4}+3t^{2}um_{t}^{4}+t^{4}m_{t}^{2}
\nn
\\
&& -2tu^{3}m_{t}^{2}-3t^{2}u^{2}m_{t}^{2}+4t^{3}um_{t}^{2}-2t^{2}u^{3}-2t^{3}u^{2},
\nn
\\
C_{5} & = & 7m_{t}^{8}-14tm_{t}^{6}-6um_{t}^{6}+10t^{2}m_{t}^{4}+2u^{2}m_{t}^{4}+10tum_{t}^{4}-2t^{3}m_{t}^{2}-4tu^{2}m_{t}^{2}-6t^{2}um_{t}^{2}+3t^{2}u^{2},
\nn
\\
C_{6} & = & 2m_{t}^{4}-tm_{t}^{2}+um_{t}^{2}-2tu.
\end{eqnarray}
We have
\begin{eqnarray}
B_{2}^{fin} & = & \frac{C_{2}}{4K^{2}s}\ln\left(-\frac{s}{\mu^{2}}\right)\left(\ln\left(-\frac{s}{\mu^{2}}\right)-4\ln\left(\frac{t_{1}}{\mu^{2}}\right)+2\ln\left(\frac{m_{t}^{2}}{\mu^{2}}\right)\right)+\frac{C_{3}m_{t}^{2}}{4K^{2}}\left(2\ln^{2}\left(\frac{t_{1}}{\mu^{2}}\right)\right.
\nn
\\
&&\left.-\ln^{2}\left(\frac{m_{t}^{2}}{\mu^{2}}\right)-4\text{Li}_{2}\left(-\frac{t}{t_{1}}\right)\right)+\frac{C_{4}}{K^{2}s_{1}s\beta}\left(\text{Li}_{2}\left(\frac{1}{1-\beta}\right)-2\text{Li}_{2}(1-\beta)+\text{Li}_{2}\left(\frac{1}{\text{\ensuremath{\beta}}+1}\right)\right.
\nn
\\
&&\left.-\text{Li}_{2}\left(\frac{\beta+1}{1-\beta}\right)+\frac{1}{2}\text{Li}_{2}\left(\frac{4m_{t}^{2}}{s}\right)-\frac{1}{2}\text{Li}_{2}\left(\frac{s}{4m_{t}^{2}}\right)+\frac{\pi^{2}}{6}\right)+\frac{\pi^{2}C_{5}}{6K^{2}s}+\frac{C_{6}\ln\left(\frac{m_{t}^{2}}{\mu^{2}}\right)m_{t}^{2}}{Ks_{1}t}
\nn
\\
&&+\frac{m_{t}^{2}}{Ks_{1}t}\left(t(u-t)\ln\left(-\frac{s}{\mu^{2}}\right)+s_{1}t_{1}\ln\left(\frac{t_{1}}{\mu^{2}}\right)\right).
\end{eqnarray}
For $B_{1}^{fin}${{,}} we define another set of coefficients
\begin{eqnarray}
C_{7} & = & 4m_{t}^{8}-9tm_{t}^{6}-um_{t}^{6}+10t{}^{2}m_{t}^{4}-5t^{3}m_{t}^{2}-t^{2}um_{t}^{2}+t^{4}+t^{2}u^{2},
\nn
\\
C_{8} & = & m_{t}^{6}-4tm_{t}^{4}+3t^{2}m_{t}^{2}-t^{3}+t^{2}u,
\nn
\\
C_{9} & = & 6sm_{t}^{8}+8tm_{t}^{8}-s^{2}m_{t}^{6}-24t^{2}m_{t}^{6}-12stm_{t}^{6}+24t^{3}m_{t}^{4}+16st^{2}m_{t}^{4}+4s^{2}tm_{t}^{4}-8t^{4}m_{t}^{2}
\nn
\\
&&-12st^{3}m_{t}^{2}-5s^{2}t^{2}m_{t}^{2}+2st^{4}+2s^{2}t^{3}+s^{3}t^{2},
\nn
\\
C_{10} & = & 7m_{t}^{8}-18tm_{t}^{6}-2um_{t}^{6}+20t^{2}m_{t}^{4}+2tum_{t}^{4}-10t^{3}m_{t}^{2}-2t^{2}um_{t}^{2}+2t^{4}+t^{2}u^{2}
\nn
\\
C_{11} & = & 2m_{t}^{4}-t^{2}-tu.
\end{eqnarray}
We have
\begin{eqnarray}
B_{1}^{fin} & = & \frac{C_{7}}{4K^{2}s}\ln\left(-\frac{s}{\mu^{2}}\right)\left(\ln\left(-\frac{s}{\mu^{2}}\right)-4\ln\left(\frac{t_{1}}{\mu^{2}}\right)+2\ln\left(\frac{m_{t}^{2}}{\mu^{2}}\right)\right)-\frac{C_{8}}{4K^{2}}\left(-2\ln^{2}\left(\frac{t_{1}}{\mu^{2}}\right)\right.
\nn
\\
&&\left.+\ln^{2}\left(\frac{m_{t}^{2}}{\mu^{2}}\right)+4\text{Li}_{2}\left(-\frac{t}{t_{1}}\right)\right)+\frac{C_{9}}{K^{2}s_{1}s\beta}\left(\text{Li}_{2}\left(\frac{1}{1-\beta}\right)-2\text{Li}_{2}(1-\beta)+\text{Li}_{2}\left(\frac{1}{\text{\ensuremath{\beta}}+1}\right)\right.
\nn
\\
&&\left.-\text{Li}_{2}\left(\frac{\beta+1}{1-\beta}\right)+\frac{1}{2}\text{Li}_{2}\left(\frac{4m_{t}^{2}}{s}\right)-\frac{1}{2}\text{Li}_{2}\left(\frac{s}{4m_{t}^{2}}\right)+\frac{\pi^{2}}{6}\right)+\frac{\pi^{2}C_{10}}{6K^{2}s}-\frac{C_{11}}{Ks_{1}}\ln\left(-\frac{s}{\mu^{2}}\right)
\nn
\\
&&+\frac{1}{Ks_{1}}\left((t-u)\ln\left(\frac{m_{t}^{2}}{\mu^{2}}\right)m_{t}^{2}-s_{1}t_{1}\ln\left(\frac{t_{1}}{\mu^{2}}\right)\right).
\end{eqnarray}
The IR-divergent parts are proportional to the tree amplitudes
\begin{eqnarray}
\mathcal{A}_{b1}\left(\lambda_{1},\lambda_{2},\lambda_{3},\lambda_{4}\right) & = & \mathcal{A}_{tree}\left(\lambda_{1},\lambda_{2},\lambda_{3},\lambda_{4}\right)\frac{\alpha_{s}}{4\pi}\left\{-\frac{2}{\epsilon_{\text{IR}}^{2}}+\frac{2}{\epsilon_{\text{IR}}}\left(2\ln\left(\frac{t_{1}}{\mu^{2}}\right)-\ln\left(\frac{m_{t}^{2}}{\mu^{2}}\right)\right)\right\}+\cdots
\nn
\\
\mathcal{A}_{b2}\left(\lambda_{1},\lambda_{2},\lambda_{3},\lambda_{4}\right) & = & \mathcal{A}_{tree}\left(\lambda_{1},\lambda_{2},\lambda_{3},\lambda_{4}\right)\frac{\alpha_{s}}{4\pi}\left\{\frac{2}{\epsilon_{\text{IR}}^{2}}-\frac{2}{\epsilon_{\text{IR}}}\left(2\ln\left(\frac{u_{1}}{\mu^{2}}\right)-\ln\left(\frac{m_{t}^{2}}{\mu^{2}}\right)\right)\right\}+\cdots
\nn
\\
\end{eqnarray}
As mentioned before, the SM results presented above agree with those in {{ref.}}~\cite{Korner:2002hy}.
\subsection{KK Gluon Induced Helicity Amplitude for $q\bar{q} \rightarrow t\bar{t}$}
\subsubsection{Results for LO Diagrams}
\begin{figure}[h!]
  \centering
    \begin{fmffile}{kktree}
    \begin{fmfgraph*}(40,25)
      \fmfleft{i2,i1}
      \fmfright{o2,o1}
      \fmf{quark}{i1,v1,i2}
      \fmf{quark}{o2,v2,o1}
      \fmf{dbl_curly}{v1,v2}
    \end{fmfgraph*}
  \end{fmffile}
  \caption{KK gluon induced tree graph for $q\bar{q} \rightarrow t\bar{t}$. Diagrams vanish identically are not shown.}
  \label{fig:kktree}
\end{figure}
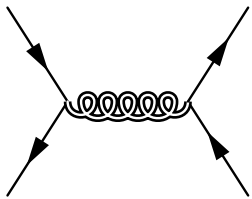
The fermionic current coupled to the massive color octet is
\begin{equation}
\bar{\psi}\gamma^{\mu}\left(C^I_L P_L+C^I_R P_R\right)\psi,
\end{equation}
where $I=q,\,t$ denotes a massless quark or massive top quark, respectively. And $P_{L,R}=(1\mp\gamma_5)/2$ are the chiral projection operators. It is straightforward to calculate tree amplitudes~(Fig.~\ref{fig:kktree}) for the KK-gluon-mediated process. The color structure is identical to that of the gluon induced diagrams, and the Lorentz part is written in terms of spinor products
\begin{eqnarray}
\mathcal{A}_{tree,KK}\left(+,-,+,+\right) & = & \frac{2 i C^q_R m_{t}}{s-m^2_{KK}}\frac{C^t_R\spa{\eta_4}.{1}\spab{\eta_3}.{\mathbf{3}}.{2}+C^t_L\spa{\eta_3}.{1}\spab{\eta_4}.{\mathbf{4}}.{2}}{\spa{\fl 3}.{\eta_3}\spa{\eta_4}.{\fl 4}},
\nn
\\
\mathcal{A}_{tree,KK}\left(-,+,+,+\right) & = & \frac{2 i C^q_L m_{t}}{s-m^2_{KK}}\frac{C^t_R\spa{\eta_4}.{2}\spab{\eta_3}.{\mathbf{3}}.{1}+C^t_L\spa{\eta_3}.{2}\spab{\eta_4}.{\mathbf{4}}.{1}}{\spa{\fl 3}.{\eta_3}\spa{\eta_4}.{\fl 4}}.
\end{eqnarray}
At the NLO, chiral magnetic-like interaction
\begin{equation}
\bar{\psi}\frac{i\sigma^{\mu\nu}q_{\nu}}{2 m_t}\left(C^I_L P_L+C^I_R P_R\right)\psi
\end{equation}
will be induced from one-loop triangle diagrams. We also provide tree amplitudes which will be used to construct one-loop amplitudes
\begin{eqnarray}
\mathcal{A}_{tree,KK}^{(m)}\left(+,-,+,+\right) & = & -\frac{i C^q_L}{m_{t}(s-m^2_{KK})}\frac{C^t_R m_{t}^{2}\spb{2}.{1}\spa{\eta_3}.{1}\spa{\eta_4}.{1}+C^t_L \spa{1}.{2}\spab{\eta_3}.{\mathbf{3}}.{2}\spab{\eta_4}.{\mathbf{4}}.{2}}{\spa{\fl 3}.{\eta_3}\spa{\eta_4}.{\fl 4}},
\nn
\\
\mathcal{A}_{tree,KK}^{(m)}\left(-,+,+,+\right) & = & \frac{i C^q_R}{m_{t}(s-m^2_{KK})}\frac{C^t_R m_{t}^{2}\spb{2}.{1}\spa{\eta_3}.{2}\spa{\eta_4}.{2}+C^t_L \spa{1}.{2}\spab{\eta_3}.{\mathbf{3}}.{1}\spab{\eta_4}.{\mathbf{4}}.{1}}{\spa{\fl 3}.{\eta_3}\spa{\eta_4}.{\fl 4}}.
\nn
\\
\end{eqnarray}

\subsubsection{Results for Self-energy Diagrams}
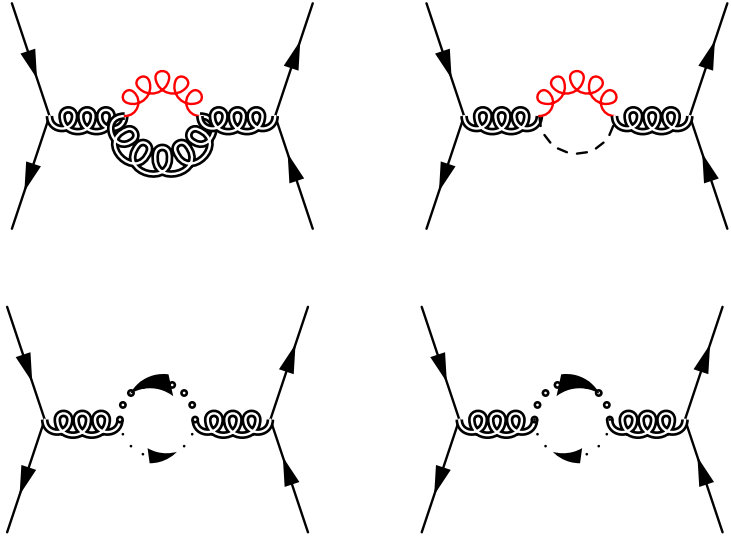
\begin{figure}[h!]
    \begin{fmffile}{kksf1}
    \begin{fmfgraph*}(50,30)
      \fmfleft{i2,i1}
      \fmfright{o2,o1}
      \fmf{quark}{i1,v1,i2}
      \fmf{quark}{o2,v4,o1}
      \fmf{phantom}{v1,v2,v3,v4}
      \fmffreeze
      \fmf{dbl_curly}{v1,v2}
      \fmf{dbl_curly}{v3,v4}
      \fmf{dbl_curly,right}{v2,v3}
      \fmf{gluon,left,tension=5,fore=red}{v2,v3}
    \end{fmfgraph*}
  \end{fmffile}
    \begin{fmffile}{kksf2}
    \begin{fmfgraph*}(50,30)
      \fmfleft{i2,i1}
      \fmfright{o2,o1}
      \fmf{quark}{i1,v1,i2}
      \fmf{quark}{o2,v4,o1}
      \fmf{phantom}{v1,v2,v3,v4}
      \fmffreeze
      \fmf{dbl_curly}{v1,v2}
      \fmf{dbl_curly}{v3,v4}
      \fmf{dashes,right}{v2,v3}
      \fmf{gluon,left,tension=5,fore=red}{v2,v3}
    \end{fmfgraph*}
  \end{fmffile}
\\
\vspace{10mm}
    \begin{fmffile}{kksf3}
    \begin{fmfgraph*}(50,30)
      \fmfleft{i2,i1}
      \fmfright{o2,o1}
      \fmf{quark}{i1,v1,i2}
      \fmf{quark}{o2,v4,o1}
      \fmf{phantom}{v1,v2,v3,v4}
      \fmffreeze
      \fmf{dbl_curly}{v1,v2}
      \fmf{dbl_curly}{v3,v4}
      \fmf{dots_arrow,right}{v2,v3}
      \fmf{dbl_dots_arrow,right,tension=5}{v3,v2}
    \end{fmfgraph*}
  \end{fmffile}
    \begin{fmffile}{kksf4}
    \begin{fmfgraph*}(50,30)
      \fmfleft{i2,i1}
      \fmfright{o2,o1}
      \fmf{quark}{i1,v1,i2}
      \fmf{quark}{o2,v4,o1}
      \fmf{phantom}{v1,v2,v3,v4}
      \fmffreeze
      \fmf{dbl_curly}{v1,v2}
      \fmf{dbl_curly}{v3,v4}
      \fmf{dbl_dots_arrow,left}{v2,v3}
      \fmf{dots_arrow,left}{v3,v2}
    \end{fmfgraph*}
  \end{fmffile}
  \caption{KK gluon induced one-loop self energy graphs for $q\bar{q} \rightarrow t\bar{t}$. Diagrams vanish identically are not shown.}
  \label{fig:kksf}
\end{figure}
At the NLO in QCD coupling $\alpha_s$, the massive KK gluon propagator~(Fig:~\ref{fig:kksf}) will receive corrections from loop of gauge bosons, their ghosts and $A^{(1)}_5$. There will be two Lorentz tensor structures $g^{\mu\nu}$ and $q^{\mu}q^{\nu}$, but the latter does not contribute to the amplitude by means of both vector-current and axial-current conservation on the massless quark side.

The color structure is identical to the tree amplitudes. We have found for the Lorentz part
\begin{eqnarray}
&&\mathcal{A}_{sf,KK}\left(\lambda_{1},\lambda_{2},\lambda_{3},\lambda_{4}\right) = \mathcal{A}_{tree,KK}\left(\lambda_{1},\lambda_{2},\lambda_{3},\lambda_{4}\right)\frac{\alpha_{s}}{\pi}\frac{s}{s-m^2_{KK}}\left\{\frac{1}{\epsilon _{\text{UV}}}\left(\frac{9 m_{KK}^2}{4 s}+\frac{5}{2}\right)\right.
\nn
\\
&&\left.+\frac{1}{12 s^3}\left[3 \ln \left(\frac{m_{KK}^2}{\mu ^2}\right) \left(2 m_{KK}^4-6 s m_{KK}^2-15 s^2\right) m_{KK}^2\right.\right.
\nn
\\
&&\left.\left.-6 \ln \left(\frac{m_{KK}^2-s}{\mu ^2}\right) \left(m_{KK}^6-3 s m_{KK}^4-3 s^2 m_{KK}^2+5 s^3\right)+s \left(-6 m_{KK}^4+51 s m_{KK}^2+56 s^2\right)\right]\right\}.
\nn
\\
\end{eqnarray}
We subtract the one-loop KK gluon propagator on the mass shell, and obtain mass renormalization and wave-function renormalization{{:}}
\begin{eqnarray}
\delta m^2_{KK} & = & m^2_{KK}\frac{\alpha_s}{\pi}\left\{-\frac{19}{4\epsilon _{\text{UV}}}+\frac{19}{4}\ln \left(\frac{m_{KK}^2}{\mu ^2}\right)-\frac{101}{12}\right\},
\nn
\\
\delta Z_{KK} & = & \frac{\alpha_s}{\pi}\left\{\frac{5}{2 \epsilon_{\text{UV}}}-\frac{3}{2 \epsilon_{\text{IR}}}-\ln \left(\frac{m_{KK}^2}{\mu ^2}\right)+\frac{13}{6}\right\}.
\end{eqnarray}
We choose $\overline{\mathrm{MS}}$ scheme to renormalize the coupling between quarks and the massive KK gluon. The counter-term contribution that cancels the UV-divergent part of KK gluon self-energy is given by
\begin{eqnarray}
\mathcal{A}_{sf,KK}^{CT}\left(\lambda_{1},\lambda_{2},\lambda_{3},\lambda_{4}\right) &=& \mathcal{A}_{tree,KK}\left(\lambda_{1},\lambda_{2},\lambda_{3},\lambda_{4}\right)\frac{\alpha_{s}}{\pi}\left\{-\delta Z^{\overline{\mathrm{MS}}}_{KK}+\frac{\delta m^2_{KK}}{s-m^2_{KK}}\right\}
\nn
\\
& = & \mathcal{A}_{tree,KK}\left(\lambda_{1},\lambda_{2},\lambda_{3},\lambda_{4}\right)\frac{\alpha_{s}}{\pi}\left\{-\frac{5}{2 \epsilon _{\text{UV}}}+\frac{5}{2}\ln \left(\frac{m_{KK}^2}{\mu ^2}\right)\right.
\nn
\\
&&\left.+\frac{m^2_{KK}}{s-m^2_{KK}}\left(-\frac{19}{4 \epsilon _{\text{UV}}}+\frac{19}{4} \ln \left(\frac{m_{KK}^2}{\mu ^2}\right)-\frac{101}{12}\right)\right\}.
\nn
\end{eqnarray}
{{Here we have include a logarithmic term $\frac{5}{2}\ln \left(\frac{m_{KK}^2}{\mu ^2}\right)$ in the definition of $\delta Z^{\overline{\mathrm{MS}}}_{KK}$,}} {{$\delta Z^{\overline{\mathrm{MS}}}_{KK} = \frac{\als}{\pi}\left( \frac{5}{2\ep_{\rm UV}} - \frac{5}{2}\ln \frac{m_{KK}^2}{\mu ^2}\right)$.}}

\subsubsection{Results for Triangle Diagrams}
\begin{figure}[h!]
    \begin{fmffile}{kkvt1}
    \begin{fmfgraph*}(50,30)
      \fmfleft{i2,i1}
      \fmfright{o2,o1}
      \fmf{quark}{i1,va,v1,vb,i2}
      \fmf{phantom}{o2,vc,v2,vd,o1}
      \fmf{dbl_curly,tension=0.3}{v1,v2}
      \fmffreeze
      \fmf{quark}{o2,v2,o1}
      \fmf{gluon,fore=red}{va,vb}
    \end{fmfgraph*}
  \end{fmffile}
    \begin{fmffile}{kkvt2}
    \begin{fmfgraph*}(50,30)
      \fmfleft{i2,i1}
      \fmfright{o2,o1}
      \fmf{phantom}{i1,va,v1,vb,i2}
      \fmf{phantom}{o2,vc,v2,vd,o1}
      \fmf{dbl_curly}{v1,v2}
      \fmffreeze
      \fmf{quark}{i1,va,vb,i2}
      \fmf{gluon,fore=red}{va,v1}
      \fmf{dbl_curly}{v1,vb}
      \fmf{quark}{o2,v2,o1}
    \end{fmfgraph*}
  \end{fmffile}
    \begin{fmffile}{kkvt3}
    \begin{fmfgraph*}(50,30)
      \fmfleft{i2,i1}
      \fmfright{o2,o1}
      \fmf{phantom}{i1,va,v1,vb,i2}
      \fmf{phantom}{o2,vc,v2,vd,o1}
      \fmf{dbl_curly}{v1,v2}
      \fmffreeze
      \fmf{quark}{i1,va,vb,i2}
      \fmf{dbl_curly}{va,v1}
      \fmf{gluon,fore=red}{v1,vb}
      \fmf{quark}{o2,v2,o1}
    \end{fmfgraph*}
  \end{fmffile}
\\
\vspace{10mm}
    \begin{fmffile}{kkvt4}
    \begin{fmfgraph*}(50,30)
      \fmfleft{i2,i1}
      \fmfright{o2,o1}
      \fmf{quark}{i1,va,v1,vb,i2}
      \fmf{phantom}{o2,vc,v2,vd,o1}
      \fmf{dbl_curly,tension=0.3}{v1,v2}
      \fmffreeze
      \fmf{quark}{o2,v2,o1}
      \fmf{gluon,fore=red}{vc,vd}
    \end{fmfgraph*}
  \end{fmffile}
    \begin{fmffile}{kkvt5}
    \begin{fmfgraph*}(50,30)
      \fmfleft{i2,i1}
      \fmfright{o2,o1}
      \fmf{phantom}{i1,va,v1,vb,i2}
      \fmf{phantom}{o2,vc,v2,vd,o1}
      \fmf{dbl_curly}{v1,v2}
      \fmffreeze
      \fmf{quark}{o2,vc,vd,o1}
      \fmf{gluon,fore=red}{vc,v2}
      \fmf{dbl_curly}{v2,vd}
      \fmf{quark}{i1,v1,i2}
    \end{fmfgraph*}
  \end{fmffile}
    \begin{fmffile}{kkvt6}
    \begin{fmfgraph*}(50,30)
      \fmfleft{i2,i1}
      \fmfright{o2,o1}
      \fmf{phantom}{i1,va,v1,vb,i2}
      \fmf{phantom}{o2,vc,v2,vd,o1}
      \fmf{dbl_curly}{v1,v2}
      \fmffreeze
      \fmf{quark}{o2,vc,vd,o1}
      \fmf{dbl_curly}{vc,v2}
      \fmf{gluon,fore=red}{v2,vd}
      \fmf{quark}{i1,v1,i2}
    \end{fmfgraph*}
  \end{fmffile}
    \begin{fmffile}{kkvt7}
    \begin{fmfgraph*}(50,30)
      \fmfleft{i2,i1}
      \fmfright{o2,o1}
      \fmf{phantom}{i1,va,v1,vb,i2}
      \fmf{phantom}{o2,vc,v2,vd,o1}
      \fmf{dbl_curly}{v1,v2}
      \fmffreeze
      \fmf{quark}{o2,vc,vd,o1}
      \fmf{dashes}{vc,v2}
      \fmf{gluon,fore=red}{v2,vd}
      \fmf{quark}{i1,v1,i2}
    \end{fmfgraph*}
  \end{fmffile}
    \begin{fmffile}{kkvt8}
    \begin{fmfgraph*}(50,30)
      \fmfleft{i2,i1}
      \fmfright{o2,o1}
      \fmf{phantom}{i1,va,v1,vb,i2}
      \fmf{phantom}{o2,vc,v2,vd,o1}
      \fmf{dbl_curly}{v1,v2}
      \fmffreeze
      \fmf{quark}{o2,vc,vd,o1}
      \fmf{gluon,fore=red}{vc,v2}
      \fmf{dashes}{v2,vd}
      \fmf{quark}{i1,v1,i2}
    \end{fmfgraph*}
  \end{fmffile}
  \caption{KK gluon induced one-loop triangle graphs for $q\bar{q} \rightarrow t\bar{t}$. Diagrams vanish identically are not shown.}
  \label{fig:kkvt}
\end{figure}
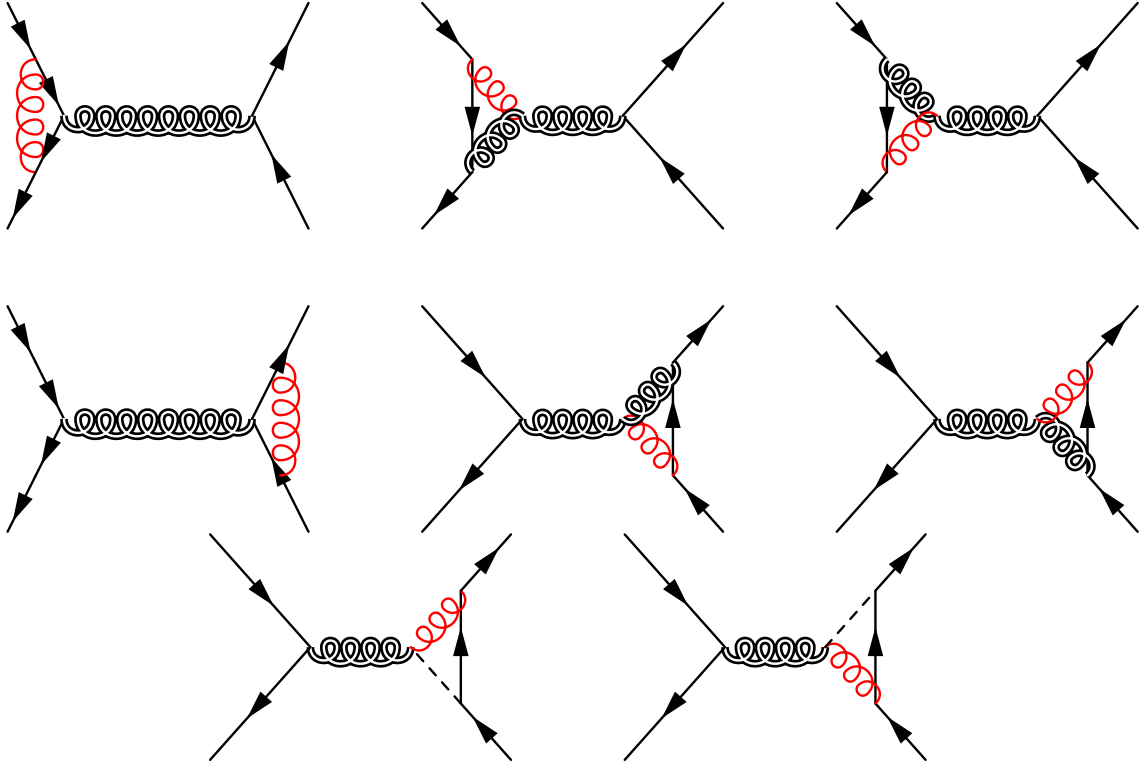
For the one-loop massless triangle diagrams~(Fig.~\ref{fig:kkvt}), the Lorentz amplitude is again the tree amplitude multiplied by a form factor. The form factor is the same for both helicity configurations {{$(+,-,+,+)$ and $(-,+,+,+)$}}. The explicit result is
\begin{eqnarray}
&&\mathcal{A}_{vt,KK}^{q}\left(\lambda_{1},\lambda_{2},\lambda_{3},\lambda_{4}\right) = \mathcal{A}_{tree,KK}\left(\lambda_{1},\lambda_{2},\lambda_{3},\lambda_{4}\right)\frac{\alpha_{s}}{\pi}\left\{\frac{53}{24 \epsilon _{\text{UV}}}+\frac{1}{12 \epsilon _{\text{IR}}^2}\right.
\nn
\\
&&\left.+\frac{1}{\epsilon _{\text{IR}}}\left[-\frac{3 m_{KK}^2}{2 s} \ln \left(\frac{m_{KK}^2-s}{m_{KK}^2}\right)-\frac{1}{12} \ln \left(-\frac{s}{\mu ^2}\right)-\frac{4}{3}\right]+\frac{1}{24} \ln ^2\left(-\frac{s}{\mu ^2}\right)-\frac{1}{8} \ln \left(-\frac{s}{\mu ^2}\right)\right.
\nn
\\
&&\left.+\frac{3}{2} C_0\left(0,0,s,m_{KK}^2,0,0\right) m_{KK}^2-\frac{3 m_{KK}^2}{4 s}+\ln \left(\frac{m_{KK}^2-s}{\mu ^2}\right) \left(\frac{3}{4}-\frac{3 m_{KK}^4}{4 s^2}\right)\right.
\nn
\\
&&\left.+\frac{3}{4} \ln \left(\frac{m_{KK}^2}{\mu ^2}\right) \left(\frac{m_{KK}^4}{s^2}-2\right)+\frac{25}{24}\right\},
\end{eqnarray}
where the one-loop scalar integrals $C_0$~\cite{Ellis:2007qk} should be understood as only retaining the finite part. The results for one-loop massive quark vertex amplitudes can be expressed as linear combination of tree amplitudes, of both helicity configurations. They have the form
\begin{eqnarray}
\mathcal{A}_{vt,KK}^{t}\left(+,-,+,+\right) & = & \frac{2 i C^q_R m_{t}}{s-m^2_{KK}}\frac{\alpha_{s}}{\pi}\frac{1}{\spa{\fl 3}.{\eta_3}\spa{\eta_4}.{\fl 4}}\left\{B^{KK}_1\left(C^t_R\spa{\eta_4}.{1}\spab{\eta_3}.{\mathbf{3}}.{2}+C^t_L\spa{\eta_3}.{1}\spab{\eta_4}.{\mathbf{4}}.{2}\right)\right.
\nn
\\
&&\left.+B^{KK}_2\left(C^t_R+C^t_L\right)\left(\spa{\eta_4}.{1}\spab{\eta_3}.{\mathbf{3}}.{2}+\spa{\eta_3}.{1}\spab{\eta_4}.{\mathbf{4}}.{2}\right)\right.
\nn
\\
&&\left.+B^{KK}_3\left(C^t_R+C^t_L\right)\left(m_{t}^{2}\spb{2}.{1}\spa{\eta_3}.{1}\spa{\eta_4}.{1}+\spa{1}.{2}\spab{\eta_3}.{\mathbf{3}}.{2}\spab{\eta_4}.{\mathbf{4}}.{2}\right)\right\},
\nn
\\
\newline
\newline
\mathcal{A}_{vt,KK}^{t}\left(-,+,+,+\right) & = & \frac{2 i C^q_L m_{t}}{s-m^2_{KK}}\frac{\alpha_{s}}{\pi}\frac{1}{\spa{\fl 3}.{\eta_3}\spa{\eta_4}.{\fl 4}}\left\{B^{KK}_1\left(C^t_R\spa{\eta_4}.{2}\spab{\eta_3}.{\mathbf{3}}.{1}+C^t_L\spa{\eta_3}.{2}\spab{\eta_4}.{\mathbf{4}}.{1}\right)\right.
\nn
\\
&&\left.+B^{KK}_2\left(C^t_R+C^t_L\right)\left(\spa{\eta_4}.{2}\spab{\eta_3}.{\mathbf{3}}.{1}+\spa{\eta_3}.{2}\spab{\eta_4}.{\mathbf{4}}.{1}\right)\right.
\nn
\\
&&\left.-B^{KK}_3\left(C^t_R+C^t_L\right)\left(m_{t}^{2}\spb{2}.{1}\spa{\eta_3}.{2}\spa{\eta_4}.{2}+\spa{1}.{2}\spab{\eta_3}.{\mathbf{3}}.{1}\spab{\eta_4}.{\mathbf{4}}.{1}\right)\right\},
\end{eqnarray}
where $B^{KK}_i,\,i=1,2,3$ are coefficients that depend on $s$, $m^2_t$ and $m^2_{KK}$. We give explicit expressions for these coefficients. Only $B^{KK}_1$ has divergent parts. The other 2 coefficients are finite. The first coefficient is
\begin{eqnarray}
&&B^{KK}_1 = \frac{53}{24 \epsilon _{\text{UV}}}+\left(\frac{2m^2_t}{s}-1\right)\frac{1}{12 \beta  \epsilon _{\text{IR}}} \ln \left(\frac{\beta +1}{\beta -1}\right)+\frac{C_1^{KK}}{24 s s_1}+\frac{s-2 m_t^2}{12} C_0\left(m_t^2,m_t^2,s,m_t^2,0,m_t^2\right)
\nn
\\
&&+\frac{3 C_2^{KK}}{2 s s_1}C_0\left(m_t^2,m_t^2,s,m_{KK}^2,m_t^2,0\right) +\frac{C_3^{KK}}{6 s s_1}\ln \left(\frac{m_t^2}{\mu ^2}\right) +\frac{3\left(m^2_{KK}+8m^2_t+s\right) m_{KK}^2}{4 s s_1} \ln \left(\frac{m_{KK}^2}{\mu ^2}\right)
\nn
\\
&&+\frac{3\left(m^2_{KK}+8m^2_t+s\right) \left(s-m_{KK}^2\right)}{4 s s_1} \ln \left(\frac{m_{KK}^2-s}{\mu ^2}\right) -\frac{3\left(2 m_{KK}^2 m_t^{2}-m_{KK}^2 s+2m_t^2 s-2s^2\right)}{4 s s_1}
\nn
\\
&&\times B_0\left(m_t^2,m_{KK}^2,m_t^2\right) -\frac{\left(4 m_t^2-3 s\right)}{24 s_1}B_0\left(s,m_t^2,m_t^2\right),
\end{eqnarray}
with polynomial coefficients
\begin{eqnarray}
C^{KK}_1 &=&47 s^2-36 m_{KK}^2 s-420 m_t^2 s+72 m_{KK}^2 m_t^2,
\nn
\\
C^{KK}_2 &=&m_t^2 m_{KK}^4+s^2 m_{KK}^2-2 s m_t^2 m_{KK}^2-3 s^2 m_t^2,
\nn
\\
C^{KK}_3 &=&-8 s^2+7 m_t^2 s-9 m_{KK}^2 m_t^2.
\end{eqnarray}
The second coefficient is
\begin{eqnarray}
&&B^{KK}_2 =-\frac{C_4^{KK}}{12 s s_1^2}-\frac{m_t^2}{12 s_1}B_0\left(s,m_t^2,m_t^2\right) +\frac{3 C_5^{KK}}{4 s s_1^2} B_0\left(m_t^2,m_{KK}^2,m_t^2\right)+\frac{3  C_8^{KK} m_{KK}^2}{4 s s_1^2}\ln \left(\frac{m_{KK}^2}{\mu ^2}\right)
\nn
\\
&&-\frac{C_7^{KK} m_t^2}{12 s s_1^2}\ln \left(\frac{m_t^2}{\mu ^2}\right) -\frac{3C_6^{KK} \left(s-m_{KK}^2\right) m_t^2}{2 s s_1^2} C_0\left(m_t^2,m_t^2,s,m_{KK}^2,m_t^2,0\right)
\nn
\\
&&+\frac{3\left(s-m_{KK}^2\right) m_t^2 \left(3 m_{KK}^2+8 m_t^2-5 s\right)}{2 s s_1^2} \ln \left(\frac{m_{KK}^2-s}{\mu ^2}\right),
\end{eqnarray}
with polynomial coefficients
\begin{eqnarray}
C^{KK}_4 &=&-72 m_{KK}^2 m_t^4+224 s m_t^4-110 s^2 m_t^2+108 s m_{KK}^2 m_t^2-9 s^2 m_{KK}^2,
\nn
\\
C^{KK}_5 &=&-4 m_{KK}^2 m_t^4-4 s^2 m_t^2+4m_t^4 s+8 s m_{KK}^2 m_t^2-s^2 m_{KK}^2,
\nn
\\
C^{KK}_6 &=&2m_t^2 s-2 s^2+m_{KK}^2 s+2 m_{KK}^2 m_t^2,
\nn
\\
C^{KK}_7 &=&-53 s^2+18 m_{KK}^2 s+104 m_t^2 s+36 m_{KK}^2 m_t^2,
\nn
\\
C^{KK}_8 &=&16 m_t^4+6 m_{KK}^2 m_t^2-6 s m_t^2-s^2.
\end{eqnarray}
The third coefficient is
\begin{eqnarray}
&&B^{KK}_3 =\frac{C_9^{KK}}{24 s m_t^2 s_1^2}+\frac{1}{24 s_1}B_0\left(s,m_t^2,m_t^2\right)+\frac{C_{12}^{KK}}{24 s s_1^2}\ln \left(\frac{m_t^2}{\mu ^2}\right)+\frac{3C_{11}^{KK}\left(s-m_{KK}^2\right)}{4 s s_1^2}
\nn
\\
&&\times C_0\left(m_t^2,m_t^2,s,m_{KK}^2,m_t^2,0\right) -\frac{3\left(s-m_{KK}^2\right) \left(-3 m_{KK}^2+8 m_t^2+s\right)}{4 s s_1^2} \ln \left(\frac{m_{KK}^2-s}{\mu ^2}\right)
\nn
\\
&&-\frac{3 C_{10}^{KK}}{8 s m_t^2 s_1^2}B_0\left(m_t^2,m_{KK}^2,m_t^2\right)-\frac{3 C_{13}^{KK} m_{KK}^2}{8 s m_t^2 s_1^2} \ln \left(\frac{m_{KK}^2}{\mu ^2}\right),
\end{eqnarray}
with polynomial coefficients
\begin{eqnarray}
C^{KK}_{9} &=&72 m_{KK}^2 m_t^4+80 s m_t^4+34 s^2 m_t^2-108 s m_{KK}^2 m_t^2+9 s^2 m_{KK}^2,
\nn
\\
C^{KK}_{10} &=&4 m_{KK}^2 m_t^4+12 s m_t^4-8 s m_{KK}^2 m_t^2+s^2 m_{KK}^2,
\nn
\\
C^{KK}_{11} &=&-2 m_t^2 m_{KK}^2-s m_{KK}^2+6 s m_t^2,
\nn
\\
C^{KK}_{12} &=&19 s^2-18 m_{KK}^2 s+32 m_t^2 s-36 m_{KK}^2 m_t^2,
\nn
\\
C^{KK}_{13} &=&16 m_t^4-6 m_{KK}^2 m_t^2-2 s m_t^2+s^2.
\end{eqnarray}
Here the scalar one-loop integrals $B_0$'s and $C_0$'s~\cite{Ellis:2007qk} should be understood as only retaining the finite part.

The renormailzation of the above two one-loop vertex amplitudes are very similar to the case of the SM process. We choose to renormalize the chiral coupling between the KK gluon and quarks in the $\overline{\mathrm{MS}}$ scheme
\begin{equation}
\label{eq:cqt}
\delta Z_{C^{q/t}_{L,R}}=-\delta Z^{\overline{\mathrm{MS}}}_{\Gamma_{KK}}-\delta Z^{\overline{\mathrm{MS}}}_{q/t}-\frac{1}{2}\delta Z^{\overline{\mathrm{MS}}}_{KK},
\end{equation}
where {{$\delta Z^{\overline{\mathrm{MS}}}_{\Gamma_{KK}}=\frac{\als}{\pi}\frac{53}{24\ep_{\rm UV}}$}} is the UV-divergent part of the one-loop vertex function, which is common regardless of the quark mass, and also regardless of the chirality of the coupling. And $\delta Z^{\overline{\mathrm{MS}}}_{q/t}$ is just the UV-divergent part of the on-shell wave-function renormalization constant for massless or massive quark.

The counter-term contributions are easily obtained:
\begin{eqnarray}
\mathcal{A}_{vt,KK}^{q,CT}\left(\lambda_{1},\lambda_{2},\lambda_{3},\lambda_{4}\right) & = & \mathcal{A}_{tree,KK}\left(\lambda_{1},\lambda_{2},\lambda_{3},\lambda_{4}\right)\times\left(-\delta Z^{\overline{\mathrm{MS}}}_{\Gamma_{KK}}-\delta Z^{\overline{\mathrm{MS}}}_q + \delta Z^{\mathrm{OS}}_q \right)
\nn
\\
& = & \mathcal{A}_{tree,KK}\left(\lambda_{1},\lambda_{2},\lambda_{3},\lambda_{4}\right)\frac{\alpha_{s}}{\pi}\left\{-\frac{53}{24\epsilon_{\text{UV}}}+\frac{1}{3\epsilon_{\text{IR}}}\right\},
\nn
\\
\mathcal{A}_{vt,KK}^{t,CT}\left(\lambda_{1},\lambda_{2},\lambda_{3},\lambda_{4}\right) & = & \mathcal{A}_{tree,KK}\left(\lambda_{1},\lambda_{2},\lambda_{3},\lambda_{4}\right)\times\left(-\delta Z^{\overline{\mathrm{MS}}}_{\Gamma_{KK}}-\delta Z^{\overline{\mathrm{MS}}}_t + \delta Z^{\mathrm{OS}}_t \right)
\nn
\\
& = & \mathcal{A}_{tree,KK}\left(\lambda_{1},\lambda_{2},\lambda_{3},\lambda_{4}\right)\frac{\alpha_{s}}{\pi}\left\{-\frac{53}{24\epsilon_{\text{UV}}}-\frac{2}{3\epsilon_{\text{IR}}}+\ln \left(\frac{m_t^2}{\mu ^2}\right)-\frac{5}{3}\right\}.
\nn
\\
\end{eqnarray}

\subsubsection{Results for Box Diagrams}
\begin{figure}[h!]
    \begin{fmffile}{kkbox1}
    \begin{fmfgraph*}(50,30)
      \fmfleft{i2,i1}
      \fmfright{o2,o1}
      \fmf{quark}{i1,v1,v2,i2}
      \fmf{quark}{o2,v4,v3,o1}
      \fmf{dbl_curly,tension=0.5}{v1,v3}
      \fmf{gluon,tension=0.5,fore=red}{v2,v4}
    \end{fmfgraph*}
  \end{fmffile}
    \begin{fmffile}{kkbox2}
    \begin{fmfgraph*}(50,30)
      \fmfleft{i2,i1}
      \fmfright{o2,o1}
      \fmf{quark}{i1,v1,v2,i2}
      \fmf{phantom}{o2,v4,v3,o1}
      \fmf{dbl_curly,tension=0.5}{v1,v3}
      \fmf{gluon,tension=0.5,fore=red}{v2,v4}
      \fmffreeze
      \fmf{quark}{o2,v3,v4,o1}
    \end{fmfgraph*}
  \end{fmffile}
\\
\vspace{10mm}
    \begin{fmffile}{kkbox3}
    \begin{fmfgraph*}(50,30)
      \fmfleft{i2,i1}
      \fmfright{o2,o1}
      \fmf{quark}{i1,v1,v2,i2}
      \fmf{quark}{o2,v4,v3,o1}
      \fmf{gluon,tension=0.5,fore=red}{v1,v3}
      \fmf{dbl_curly,tension=0.5}{v2,v4}
    \end{fmfgraph*}
  \end{fmffile}
    \begin{fmffile}{kkbox4}
    \begin{fmfgraph*}(50,30)
      \fmfleft{i2,i1}
      \fmfright{o2,o1}
      \fmf{quark}{i1,v1,v2,i2}
      \fmf{phantom}{o2,v4,v3,o1}
      \fmf{gluon,tension=0.5,fore=red}{v1,v3}
      \fmf{dbl_curly,tension=0.5}{v2,v4}
      \fmffreeze
      \fmf{quark}{o2,v3,v4,o1}
    \end{fmfgraph*}
  \end{fmffile}
  \caption{KK gluon induced regular and cross box diagrams. Diagrams vanish identically are not shown.}
  \label{fig:kkbox}
\end{figure}
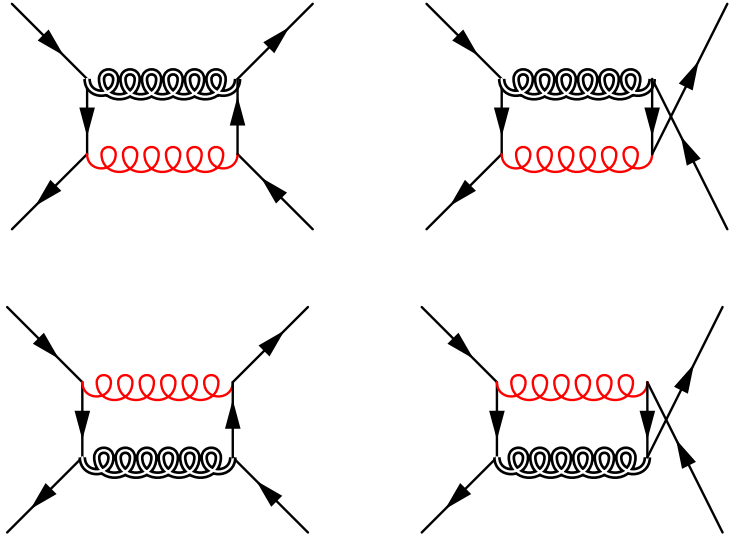
For the KK gluon-mediated process at one-loop, there are 2 regular box diagrams and 2 crossed box diagrams~(Fig.~\ref{fig:kkbox}). We do not repeat the color structure which is identical to that of the Standard Model box diagrams. The Lorentz part can be expressed in terms of the same set of spinor products basis combined with both left- and right-handed couplings, with a total of 6 coefficients $B^{KK}_{i,1/2},\,i=4,5,6$ which depend on $s,t,u,m^2_t,m^2_{KK}$. For the regular box diagrams with both helicity configurations, the amplitudes are written as
\begin{eqnarray}
\mathcal{A}_{b1,KK}\left(+,-,+,+\right) & = & \frac{2 i C^q_R m_{t}}{\spa{\fl 3}.{\eta_3}\spa{\eta_4}.{\fl 4}}\frac{\alpha_{s}}{\pi}\left\{\left(B^{KK}_{4,1}C^t_R\spa{\eta_4}.{1}\spab{\eta_3}.{\mathbf{3}}.{2}+B^{KK}_{4,2}C^t_L\spa{\eta_3}.{1}\spab{\eta_4}.{\mathbf{4}}.{2}\right)\right.
\nn
\\
&&\left.+\left(B^{KK}_{5,1}C^t_L\spa{\eta_4}.{1}\spab{\eta_3}.{\mathbf{3}}.{2}+B^{KK}_{5,2}C^t_R\spa{\eta_3}.{1}\spab{\eta_4}.{\mathbf{4}}.{2}\right)\right.
\nn
\\
&&\left.+\left(B^{KK}_{6,1}C^t_R+B^{KK}_{6,2}C^t_L\right)\left(m_{t}^{2}\spb{2}.{1}\spa{\eta_3}.{1}\spa{\eta_4}.{1}+\spa{1}.{2}\spab{\eta_3}.{\mathbf{3}}.{2}\spab{\eta_4}.{\mathbf{4}}.{2}\right)\right\},
\nn
\\
\mathcal{A}_{b1,KK}\left(-,+,+,+\right) & = & \frac{2 i C^q_L m_{t}}{\spa{\fl 3}.{\eta_3}\spa{\eta_4}.{\fl 4}}\frac{\alpha_{s}}{\pi}\left\{\left(B^{KK}_{4,2}C^t_R\spa{\eta_4}.{2}\spab{\eta_3}.{\mathbf{3}}.{1}+B^{KK}_{4,1}C^t_L\spa{\eta_3}.{2}\spab{\eta_4}.{\mathbf{4}}.{1}\right)\right.
\nn
\\
&&\left.+\left(B^{KK}_{5,2}C^t_L\spa{\eta_4}.{2}\spab{\eta_3}.{\mathbf{3}}.{1}+B^{KK}_{5,1}C^t_R\spa{\eta_3}.{2}\spab{\eta_4}.{\mathbf{4}}.{1}\right)\right.
\nn
\\
&&\left.-\left(B^{KK}_{6,2}C^t_R+B^{KK}_{6,1}C^t_L\right)\left(m_{t}^{2}\spb{2}.{1}\spa{\eta_3}.{2}\spa{\eta_4}.{2}+\spa{1}.{2}\spab{\eta_3}.{\mathbf{3}}.{1}\spab{\eta_4}.{\mathbf{4}}.{1}\right)\right\}.
\nn
\\
\end{eqnarray}
For the crossed box diagrams the amplitudes are related. We denote $\tilde{B}^{KK}_{i,1/2}$ by exchanging $t$ and $u$
\begin{equation}
\tilde{B}^{KK}_{i,1/2}=-B^{KK}_{i,1/2}(t\longleftrightarrow u),
\end{equation}
Amplitudes for crossed box diagrams are in a similar form
\begin{eqnarray}
\mathcal{A}_{b2,KK}\left(+,-,+,+\right) & = & \frac{2 i C^q_R m_{t}}{\spa{\fl 3}.{\eta_3}\spa{\eta_4}.{\fl 4}}\frac{\alpha_{s}}{\pi}\left\{\left(\tilde{B}^{KK}_{4,2}C^t_R\spa{\eta_4}.{1}\spab{\eta_3}.{\mathbf{3}}.{2}+\tilde{B}^{KK}_{4,1}C^t_L\spa{\eta_3}.{1}\spab{\eta_4}.{\mathbf{4}}.{2}\right)\right.
\nn
\\
&&\left.+\left(\tilde{B}^{KK}_{5,2}C^t_L\spa{\eta_4}.{1}\spab{\eta_3}.{\mathbf{3}}.{2}+\tilde{B}^{KK}_{5,1}C^t_R\spa{\eta_3}.{1}\spab{\eta_4}.{\mathbf{4}}.{2}\right)\right.
\nn
\\
&&\left.+\left(\tilde{B}^{KK}_{6,2}C^t_R+\tilde{B}^{KK}_{6,1}C^t_L\right)\left(m_{t}^{2}\spb{2}.{1}\spa{\eta_3}.{1}\spa{\eta_4}.{1}+\spa{1}.{2}\spab{\eta_3}.{\mathbf{3}}.{2}\spab{\eta_4}.{\mathbf{4}}.{2}\right)\right\},
\nn
\\
\mathcal{A}_{b2,KK}\left(-,+,+,+\right) & = & \frac{2 i C^q_L m_{t}}{\spa{\fl 3}.{\eta_3}\spa{\eta_4}.{\fl 4}}\frac{\alpha_{s}}{\pi}\left\{\left(\tilde{B}^{KK}_{4,1}C^t_R\spa{\eta_4}.{2}\spab{\eta_3}.{\mathbf{3}}.{1}+\tilde{B}^{KK}_{4,2}C^t_L\spa{\eta_3}.{2}\spab{\eta_4}.{\mathbf{4}}.{1}\right)\right.
\nn
\\
&&\left.+\left(\tilde{B}^{KK}_{5,1}C^t_L\spa{\eta_4}.{2}\spab{\eta_3}.{\mathbf{3}}.{1}+\tilde{B}^{KK}_{5,2}C^t_R\spa{\eta_3}.{2}\spab{\eta_4}.{\mathbf{4}}.{1}\right)\right.
\nn
\\
&&\left.-\left(\tilde{B}^{KK}_{6,1}C^t_R+\tilde{B}^{KK}_{6,2}C^t_L\right)\left(m_{t}^{2}\spb{2}.{1}\spa{\eta_3}.{2}\spa{\eta_4}.{2}+\spa{1}.{2}\spab{\eta_3}.{\mathbf{3}}.{1}\spab{\eta_4}.{\mathbf{4}}.{1}\right)\right\}.
\nn
\\
\end{eqnarray}
There is no UV divergence, and IR divergence is proportional to the treel amplitudes
\begin{eqnarray}
\mathcal{A}_{b1,KK}\left(\lambda_{1},\lambda_{2},\lambda_{3},\lambda_{4}\right) & = & \mathcal{A}_{tree,KK}\left(\lambda_{1},\lambda_{2},\lambda_{3},\lambda_{4}\right)\frac{\alpha_{s}}{4\pi}\left\{-\frac{2}{\epsilon_{\text{IR}}^{2}}+\frac{2}{\epsilon_{\text{IR}}}\left(2\ln\left(\frac{t_{1}}{\mu^{2}}\right)\right.\right.
\nn
\\
&&\left.\left.-\ln\left(\frac{m_{t}^{2}}{\mu^{2}}\right)+\frac{2m^2_{KK}}{s}\ln\left(\frac{m^2_{KK}-s}{m^2_{KK}}\right)\right)\right\}+\cdots
\nn
\\
\mathcal{A}_{b2,KK}\left(\lambda_{1},\lambda_{2},\lambda_{3},\lambda_{4}\right) & = & \mathcal{A}_{tree,KK}\left(\lambda_{1},\lambda_{2},\lambda_{3},\lambda_{4}\right)\frac{\alpha_{s}}{4\pi}\left\{\frac{2}{\epsilon_{\text{IR}}^{2}}-\frac{2}{\epsilon_{\text{IR}}}\left(2\ln\left(\frac{u_{1}}{\mu^{2}}\right)\right.\right.
\nn
\\
&&\left.\left.-\ln\left(\frac{m_{t}^{2}}{\mu^{2}}\right)+\frac{2m^2_{KK}}{s}\ln\left(\frac{m^2_{KK}-s}{m^2_{KK}}\right)\right)\right\}+\cdots
\nn
\\
\end{eqnarray}
Next we give explicit results for coefficients $B^{KK}_{i,1/2},\,i=4,5,6$. The first two coefficients have the same IR-divergent part. The rest of them are finite. The first coefficient is
\begin{eqnarray}
B^{KK}_{4,1}&=&\frac{1}{s-m^2_{KK}}\left[-\frac{1}{2\epsilon_{\text{IR}}^{2}}+\frac{1}{\epsilon_{\text{IR}}}\left(\ln\left(\frac{t_{1}}{\mu^{2}}\right)-\frac{1}{2}\ln\left(\frac{m_{t}^{2}}{\mu^{2}}\right)+\frac{m^2_{KK}}{s}\ln\left(\frac{m^2_{KK}-s}{m^2_{KK}}\right)\right)\right]
\nn
\\
&&+\frac{C_{14}^{KK}}{2 K^2}C_0\left(0,0,s,m_{KK}^2,0,0\right) -\frac{C_{15}^{KK} t_1}{2 K^2}C_0\left(m_t^2,0,t,m_t^2,0,0\right)
\nn
\\
&&-\frac{ C_{15}^{KK} t_1}{2 K^2}C_0\left(m_t^2,0,t,m_t^2,m_{KK}^2,0\right)+\frac{C_{17}^{KK} t_1}{2 K^2}D_0\left(m_t^2,m_t^2,0,0,s,t,m_{KK}^2,m_t^2,0,0\right)
\nn
\\
&&-\frac{ t_1}{K}\ln \left(\frac{t_1}{\mu ^2}\right)+\frac{C_{16}^{KK}}{2 K^2 s_1}C_0\left(m_t^2,m_t^2,s,m_{KK}^2,m_t^2,0\right)
\nn
\\
&&-\frac{ C_{18}^{KK}}{K s_1}\ln \left(\frac{m_t^2}{\mu ^2}\right)-\frac{C_{19}^{KK} m_{KK}^2}{K s s_1}\ln \left(\frac{m_{KK}^2}{\mu ^2}\right) +\frac{C_{19}^{KK} \left(m_{KK}^2-s\right)}{K s s_1}\ln \left(\frac{m_{KK}^2-s}{\mu ^2}\right)
\nn
\\
&& -\frac{ m_t^2 \left(m_t^2+t\right)}{K s_1}B_0\left(m_t^2,m_{KK}^2,m_t^2\right)+\frac{2 m_t^2 \left(m_t^2+t\right)}{K s_1},
\nn
\\
\end{eqnarray}
with the the polynomial coefficients
\begin{eqnarray}
C^{KK}_{14}&=&4 m_t^8-9 t m_t^6-u m_t^6+10 t^2 m_t^4-5 t^3 m_t^2+s t m_{KK}^2 m_t^2-t^2 u m_t^2+t^4+t^2 u^2-s t^2 m_{KK}^2,
\nn
\\
C^{KK}_{15}&=&m_t^6-4 t m_t^4+3 t^2 m_t^2+t m_{KK}^2 m_t^2-t^3-t^2 m_{KK}^2+t^2 u,
\nn
\\
C^{KK}_{16}&=&-2 m_{KK}^2 m_t^8+6 s m_t^8+8 t m_t^8-s^2 m_t^6-24 t^2 m_t^6+2 t m_{KK}^2 m_t^6-12 s t m_t^6+24 t^3 m_t^4
\nn
\\
&&+16 s t^2 m_t^4+2 t^2 m_{KK}^2 m_t^4+2 s t m_{KK}^2 m_t^4+4 s^2 t m_t^4-8 t^4 m_t^2-12 s t^3 m_t^2-5 s^2 t^2 m_t^2
\nn
\\
&&-2 t^3 m_{KK}^2 m_t^2+2 s t^2 m_{KK}^2 m_t^2-s^2 t m_{KK}^2 m_t^2+2 s t^4+2 s^2 t^3+s^3 t^2-s^2 t^2 m_{KK}^2,
\nn
\\
C^{KK}_{17}&=&4 m_t^8-m_{KK}^2 m_t^6-9 t m_t^6-u m_t^6+10 t^2 m_t^4+6 t m_{KK}^2 m_t^4-t m_{KK}^4 m_t^2-5 t^3 m_t^2
\nn
\\
&&-6 t^2 m_{KK}^2 m_t^2-t u m_{KK}^2 m_t^2-t^2 u m_t^2+t^4+t^2 m_{KK}^4+t^2 u^2+2 t^3 m_{KK}^2,
\nn
\\
C^{KK}_{18}&=&m_t^2 \left(m_t^2+u\right),
\nn
\\
C^{KK}_{19}&=&2 m_t^4-t^2-t u.
\end{eqnarray}
The second coefficient is
\begin{eqnarray}
B^{KK}_{4,2}&=&\frac{1}{s-m^2_{KK}}\left[-\frac{1}{2\epsilon_{\text{IR}}^{2}}+\frac{1}{\epsilon_{\text{IR}}}\left(\ln\left(\frac{t_{1}}{\mu^{2}}\right)-\frac{1}{2}\ln\left(\frac{m_{t}^{2}}{\mu^{2}}\right)+\frac{m^2_{KK}}{s}\ln\left(\frac{m^2_{KK}-s}{m^2_{KK}}\right)\right)\right]
\nn
\\
&&+\frac{ t_1 m_t^2}{2 K}C_0\left(m_t^2,0,t,m_t^2,0,0\right)+\frac{t_1 m_t^2}{2 K}C_0\left(m_t^2,0,t,m_t^2,m_{KK}^2,0\right)
\nn
\\
&&+\frac{ \left(t m_t^2+u m_t^2-2 t u\right)}{2 K}C_0\left(0,0,s,m_{KK}^2,0,0\right)+\frac{2 m_t^4+t m_t^2-u m_t^2-2 t u}{2 K}
\nn
\\
&&\times C_0\left(m_t^2,m_t^2,s,m_{KK}^2,m_t^2,0\right) +\frac{ \left(m_{KK}^2 m_t^2+t m_t^2+u m_t^2-2 t u\right) t_1}{2 K}
\nn
\\
&&\times D_0\left(m_t^2,m_t^2,0,0,s,t,m_{KK}^2,m_t^2,0,0\right).
\end{eqnarray}
The third coefficient is found to be vanishing. The fourth coefficient is
\begin{eqnarray}
B^{KK}_{5,2}&=&-\frac{\left(m_{KK}^2-s\right) m_t^2 C_{20}^{KK}}{2 K^2 s_1}C_0\left(m_t^2,m_t^2,s,m_{KK}^2,m_t^2,0\right) +\frac{ \left(m_{KK}^2-s\right){}^2 m_t^2 t_1^2}{2 K^2}
\nn
\\
&&\times D_0\left(m_t^2,m_t^2,0,0,s,t,m_{KK}^2,m_t^2,0,0\right)+\frac{ \left(m_{KK}^2-s\right) m_t^2 t_1^2}{2 K^2}C_0\left(m_t^2,0,t,m_t^2,0,0\right)
\nn
\\
&&+\frac{\left(m_{KK}^2-s\right) m_t^2 t_1^2}{2 K^2}C_0\left(m_t^2,0,t,m_t^2,m_{KK}^2,0\right) +\frac{ m_t^2 t_1}{K t}\ln \left(\frac{t_1}{\mu ^2}\right)-\frac{s\left(m_{KK}^2-s\right) m_t^2 t_1}{2 K^2}
\nn
\\
&&\times C_0\left(0,0,s,m_{KK}^2,0,0\right)+\frac{ C_{21}^{KK} m_t^2}{K t s_1}\ln \left(\frac{m_t^2}{\mu ^2}\right)-\frac{(t-u) m_{KK}^2 m_t^2}{K s s_1} \ln \left(\frac{m_{KK}^2}{\mu ^2}\right)
\nn
\\
&&+\frac{(t-u)  \left(m_{KK}^2-s\right) m_t^2}{K s s_1}\ln \left(\frac{m_{KK}^2-s}{\mu ^2}\right)+\frac{ m_t^2 \left(m_t^2+u\right)}{K s_1}B_0\left(m_t^2,m_{KK}^2,m_t^2\right)
\nn
\\
&&-\frac{2 m_t^2 \left(m_t^2+u\right)}{K s_1},
\nn
\\
\end{eqnarray}
with the polynomial coefficients
\begin{eqnarray}
C^{KK}_{20}&=&2 m_t^6+4 t m_t^4-2 u m_t^4-t^2 m_t^2-u^2 m_t^2-t^3+t u^2-2 t^2 u,
\nn
\\
C^{KK}_{21}&=&2 m_t^4+u m_t^2-t u.
\end{eqnarray}
The fifth coefficient is
\begin{eqnarray}
B^{KK}_{6,1}&=&\frac{ C_{22}^{KK}}{4 K^2}C_0\left(0,0,s,m_{KK}^2,0,0\right)-\frac{ t_1 C_{22}^{KK}}{4 K^2 s}C_0\left(m_t^2,0,t,m_t^2,0,0\right)-\frac{1}{K}\ln \left(\frac{t_1}{\mu ^2}\right)
\nn
\\
&&+\frac{ C_{25}^{KK} \left(m_{KK}^2-s\right) t_1}{4 K^2 s}D_0\left(m_t^2,m_t^2,0,0,s,t,m_{KK}^2,m_t^2,0,0\right)+\frac{ C_{24}^{KK}}{4 K^2 s_1}
\nn
\\
&&\times C_0\left(m_t^2,m_t^2,s,m_{KK}^2,m_t^2,0\right)-\frac{ m_{KK}^2 \left(m_t^2+t\right)}{K s s_1}\ln \left(\frac{m_{KK}^2}{\mu ^2}\right)+\frac{ \left(m_{KK}^2-s\right) \left(m_t^2+t\right)}{K s s_1}
\nn
\\
&&\times \ln \left(\frac{m_{KK}^2-s}{\mu ^2}\right)-\frac{ C_{23}^{KK}}{4 K^2 s t_1}C_0\left(m_t^2,0,t,m_t^2,m_{KK}^2,0\right)-\frac{ C_{26}^{KK}}{2 K s_1 t_1}\ln \left(\frac{m_t^2}{\mu ^2}\right)
\nn
\\
&&-\frac{ \left(m_t^4+2 t m_t^2+t u\right)}{2 K s_1 t_1}B_0\left(m_t^2,m_{KK}^2,m_t^2\right)+\frac{m_t^4+2 t m_t^2+t u}{K s_1 t_1},
\nn
\\
\end{eqnarray}
with the polynomial coefficients
\begin{eqnarray}
C^{KK}_{22}&=&5 m_t^6-m_{KK}^2 m_t^4-11 t m_t^4-2 u m_t^4+8 t^2 m_t^2+4 t m_{KK}^2 m_t^2+3 t u m_t^2-2 t^3-2 t^2 m_{KK}^2
\nn
\\
&&-t u m_{KK}^2-t^2 u,
\nn
\\
C^{KK}_{23}&=&3 m_t^{10}+m_{KK}^2 m_t^8-19 t m_t^8-2 u m_t^8+35 t^2 m_t^6+6 t m_{KK}^2 m_t^6+11 t u m_t^6-29 t^3 m_t^4
\nn
\\
&&-11 t^2 m_{KK}^2 m_t^4-5 t u m_{KK}^2 m_t^4-13 t^2 u m_t^4+12 t^4 m_t^2-2 t^2 u^2 m_t^2+8 t^3 m_{KK}^2 m_t^2
\nn
\\
&&+2 t^2 u m_{KK}^2 m_t^2+5 t^3 u m_t^2-2 t^5+2 t^3 u^2-2 t^4 m_{KK}^2+2 t^2 u^2 m_{KK}^2-t^3 u m_{KK}^2-t^4 u,
\nn
\\
C^{KK}_{24}&=&10 m_t^8-4 m_{KK}^2 m_t^6+7 t m_t^6-3 u m_t^6-9 t^2 m_t^4-2 u^2 m_t^4-3 t m_{KK}^2 m_t^4-u m_{KK}^2 m_t^4
\nn
\\
&&-5 t u m_t^4+3 t u^2 m_t^2+4 t^2 m_{KK}^2 m_t^2-7 t^2 u m_t^2+2 t^4+t^2 u^2+2 t^3 m_{KK}^2-t u^2 m_{KK}^2
\nn
\\
&&+3 t^2 u m_{KK}^2+3 t^3 u,
\nn
\\
C^{KK}_{25}&=&-5 m_t^6+m_{KK}^2 m_t^4+11 t m_t^4+2 u m_t^4-8 t^2 m_t^2-4 t m_{KK}^2 m_t^2-3 t u m_t^2+2 t^3+2 t^2 m_{KK}^2
\nn
\\
&&+t u m_{KK}^2+t^2 u,
\nn
\\
C^{KK}_{26}&=&3 m_t^4+2 u m_t^2-t u.
\end{eqnarray}
The sixth coefficient is
\begin{eqnarray}
B^{KK}_{6,2}&=&\frac{t_1^2}{4 K s}C_0\left(m_t^2,0,t,m_t^2,0,0\right) +\frac{\left(m_{KK}^2-s\right) t_1^2}{4 K s}D_0\left(m_t^2,m_t^2,0,0,s,t,m_{KK}^2,m_t^2,0,0\right)
\nn
\\
&&-\frac{ t_1}{4 K}C_0\left(0,0,s,m_{KK}^2,0,0\right)+\frac{m_t^2+t}{4 K}C_0\left(m_t^2,m_t^2,s,m_{KK}^2,m_t^2,0\right)
\nn
\\
&&-\frac{m_t^4+2 t m_t^2-t^2-2 t u}{4 K s}C_0\left(m_t^2,0,t,m_t^2,m_{KK}^2,0\right).
\end{eqnarray}
Again, all scalar one-loop integrals should be understood as only retaining the finite part.

\section{Discussion and Conclusion}
\label{sec:conclusion}
We have presented the one-loop helicity amplitudes for $t\bar{t}$ production induced by KK gluon. The results are expressed in terms of four independent spinor products. A {{special}} feature of our calculation is that only interaction vertices that are uniquely fixed by gauge symmetry are considered{{, except}} the couplings between quark and the first KK mode, which are not fixed by gauge invariance, but are required by the LO process. Choosing the vertices this way allows our calculation to be model independent as much as possible. In fact, except for the couplings between quark and the first KK mode, the only model dependent information are the color representation and the mass of KK gluon. In this way, all the infrared QCD effects are captured in our calculation, which usually dominate fixed order cross section~\cite{Ahrens:2011px}. This implies that with minor modification, our calculation results can be applied to a variety of models containing a massive color octet, {{cf. refs.}}~\cite{Pati:1974zv,Hall:1985wz,Hewett:1988xc,Hill:1991at,Lane:1991qh,Appelquist:2000nn,Davoudiasl:1999tf,Ferrario:2009bz,Morrissey:2009tf,Chivukula:2010fk,Xiao:2010ph,Bai:2011ed,Zerwekh:2011wf,Shu:2011au,Barcelo:2011fw,Haisch:2011up}. To confirm this, we also derive the relevant Feynman rules in a model with $SU(3)_L \times SU(3)_R$ symmetry, which is spontaneously broken to diagonal $SU(3)_C$ by a bi-triplet scalar field. As expected, all Feynman rules that are {{uniquely determined}} by gauge symmetry are the same in the two models, including those vertices involving ghost and goldstone bosons~\footnote{$A^{(1)}_5$ plays the role of goldstone boson in RS model.}.

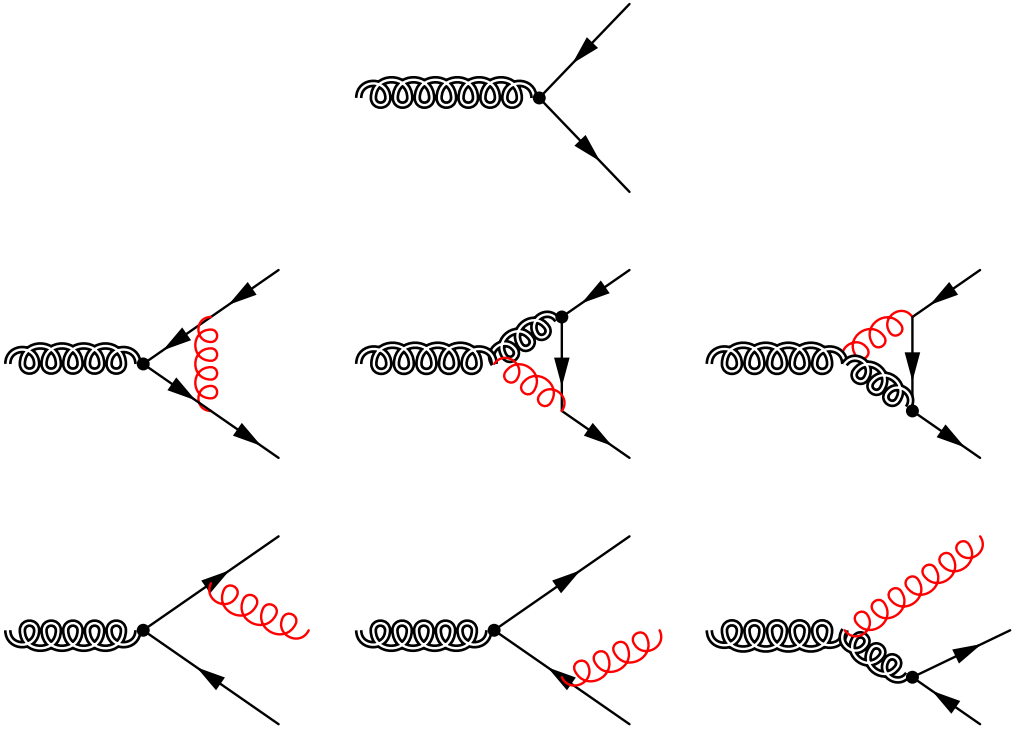
\begin{figure}[h!]
\centering
  \begin{fmffile}{lo}
    \begin{fmfgraph}(40,25)
      \fmfleft{o1}
      \fmfright{i2,i1}
      \fmf{quark}{i1,v1,i2}
      \fmf{dbl_curly}{v1,o1}
      \fmfdot{v1}
    \end{fmfgraph}
  \end{fmffile}
\\
\vspace{10mm}
    \begin{fmffile}{v1}
    \begin{fmfgraph}(40,25)
      \fmfright{i2,i1}
      \fmfleft{o1}
      \fmf{quark}{i1,v1,v2,v3,i2}
      \fmf{dbl_curly}{v2,o1}
      \fmffreeze
      \fmf{gluon,fore=red}{v1,v3}
      \fmfdot{v2}
    \end{fmfgraph}
  \end{fmffile}
    \begin{fmffile}{v2}
    \begin{fmfgraph}(40,25)
      \fmfright{i2,i1}
      \fmfleft{o1}
      \fmf{quark}{i1,v1}
      \fmf{dbl_curly}{v1,v2}
      \fmf{dbl_curly}{v2,o1}
      \fmf{quark}{v3,i2}
      \fmf{gluon,fore=red}{v3,v2}
      \fmffreeze
      \fmf{quark}{v1,v3}
      \fmfdot{v1}
    \end{fmfgraph}
  \end{fmffile}
    \begin{fmffile}{v3}
    \begin{fmfgraph}(40,25)
      \fmfright{i2,i1}
      \fmfleft{o1}
      \fmf{quark}{i1,v1}
      \fmf{gluon,fore=red}{v1,v2}
      \fmf{dbl_curly}{v2,o1}
      \fmf{quark}{v3,i2}
      \fmf{dbl_curly}{v3,v2}
      \fmffreeze
      \fmf{quark}{v1,v3}
      \fmfdot{v3}
    \end{fmfgraph}
  \end{fmffile}
\\
\vspace{10mm}
    \begin{fmffile}{v4}
    \begin{fmfgraph}(40,25)
      \fmfleft{i1}
      \fmfright{o3,o2,o1}
      \fmf{dbl_curly}{i1,v1}
      \fmf{phantom}{o3,va,v1,vb,o1}
      \fmffreeze
      \fmf{quark}{o3,v1,o1}
      \fmf{gluon,fore=red}{vb,o2}
      \fmfdot{v1}
    \end{fmfgraph}
  \end{fmffile}
    \begin{fmffile}{v5}
    \begin{fmfgraph}(40,25)
      \fmfleft{i1}
      \fmfright{o3,o2,o1}
      \fmf{dbl_curly}{i1,v1}
      \fmf{phantom}{o3,va,v1,vb,o1}
      \fmffreeze
      \fmf{quark}{o3,v1,o1}
      \fmf{gluon,fore=red}{va,o2}
      \fmfdot{v1}
    \end{fmfgraph}
  \end{fmffile}
    \begin{fmffile}{v6}
    \begin{fmfgraph}(40,25)
      \fmfleft{i1}
      \fmfright{o3,o2,o1}
      \fmf{phantom}{i1,v1}
      \fmf{phantom}{o3,va,v1,vb,o1}
      \fmffreeze
      \fmf{quark}{o3,va,o2}
      \fmf{dbl_curly}{i1,v1}
      \fmf{dbl_curly}{v1,va}
      \fmf{gluon,fore=red}{v1,o1}
      \fmfdot{va}
    \end{fmfgraph}
  \end{fmffile}
\\
\vspace{10mm}
  \caption{LO and NLO corrections to total decay width of KK gluon.}
  \label{fig:decay}
\end{figure}

Using the Feynman rules derived in this paper, the decay width of KK gluon can also be calculated to NLO. The relevant Feynman diagrams are depicted in Fig.~\ref{fig:decay}.
{{After including both virtual corrections and real emission contributions,}} the NLO decay width can be written as
\begin{equation}
\Gamma_{KK}=m_{KK}\displaystyle\sum\limits_{I=u,d,...,t}\frac{\left(C_{L}^{I}\left(\mu_{R}\right)\right)^{2}+\left(C_{R}^{I}\left(\mu_{R}^{}\right)\right)^{2}}{48\pi}\left[1+\frac{\alpha_{s}}{\pi}\left(\frac{167}{12}-\pi^{2}-\frac{15}{4}\ln\frac{m_{KK}^{2}}{\mu_{R}^{2}}\right)\right],
\end{equation}
where $I$ is a quark flavor index{{, and the corrections of $\mathcal{O}(m^2_t/m^2_{KK})$ have been neglected.}} $C^I_{L/R}(\mu_R)$ is the running coupling between quark and KK gluon. The RG-running of $C^I_{L/R}(\mu)$ can be read off from Eq.~(\ref{eq:cqt}) and is given by
\begin{equation}
  \frac{C^I_{L/R}(\mu)}{C^I_{L/R}(\mu_0)} = \left(
\frac{ \als(\mu)}{\als(\mu_0)} \right)^{15/(2\beta_0)}
\end{equation}
where $\beta_0 = 23/3$ is the QCD beta function for $N_C = 3$, $n_f=5$. For large $C^I_{L/R}$, the total decay width is large, $\sim 10\%$. This invalidates the narrow width approximation, and is one of the motivation of this work. A simple framework for dealing with virtual particles with large width is the so-called complex mass scheme~\cite{Denner:2006ic}. In this scheme, $m^2_{KK}$ is complex,
\begin{equation}
  m^2_{KK} = \tilde{m}^2_{KK} - i \tilde{m}_{KK} \Gamma_{KK},
\end{equation}
where $\tilde{m}_{KK}$ is a real mass. All the mass terms in the Feynman rules and in the helicity amplitudes should be understood as complex number.

In conclusion, we have calculated the one-loop amplitudes for
$t\bar{t}$ production induced by KK gluon. As mentioned above, the
calculation presented in this paper shows for the first time how to
calculate renormalized one-loop amplitudes {{predicted by new physics model}}. To obtain
phenomenological relevant numerical result, we need to combine
virtual and real corrections to cancel the remaining IR divergences and obtain a finite cross section. This will be presented elsewhere~\cite{hxzhu2011c}.

\begin{acknowledgments}
We would like to thank Ding Yu Shao for helpful correspondence, and Jing Shu for useful discussion. This work was supported in part by the National Natural Science Foundation of China, under Grants No.11021092 and No.10975004. C.P.Y acknowledges the support of the U.S. National Science Foundation under Grand No. PHY-0855561.
\end{acknowledgments}

\appendix

\section{Relevant Feynman Rules}
\label{sec:appendix} We collect the relevant Feynman rules that
enter our calculation in this appendix. {{All the momenta are flowing into the vertices in this section.}} The coupling between the
first KK mode and SM quark are given by
\\
\begin{equation}
\parbox{4cm}{
  \begin{fmffile}{feyn1}
    \begin{fmfgraph*}(25,25)
      \fmfleft{i2,i1}
      \fmfright{o1}
      \fmflabel{$i$}{i1}
      \fmflabel{$j$}{i2}
      \fmflabel{$\mu,a$}{o1}
      \fmf{quark}{i1,v1,i2}
      \fmf{dbl_curly}{v1,o1}
      \fmfdot{v1}
    \end{fmfgraph*}
  \end{fmffile}
}
=iT^a_{ji}\gamma^\mu \left(C^{q,t}_L \frac{1-\gamma_5}{2} + C^{q,t}_R
\frac{1+\gamma_5}{2}\right)
\end{equation}
\\
\\
\begin{equation}
\parbox{4cm}{
  \begin{fmffile}{feyn1b}
    \begin{fmfgraph*}(25,25)
      \fmfleft{i2,i1}
      \fmfright{o1}
      \fmflabel{$i$}{i1}
      \fmflabel{$j$}{i2}
      \fmflabel{$a$}{o1}
      \fmf{quark}{i1,v1,i2}
      \fmf{dashes}{v1,o1}
      \fmfdot{v1}
    \end{fmfgraph*}
  \end{fmffile}
}
= \frac{m_{q,t}}{m_{KK}} (C^{q,t}_L - C^{q,t}_R) \gamma_5 T^a_{ji}
\end{equation}
\\
The trilinear and quartic coupling between gluon and KK gluon are given by 
\\
\begin{equation}
\parbox{4cm}{
  \begin{fmffile}{feyn2}
    \begin{fmfgraph*}(25,25)
      \fmfleft{i2,i1}
      \fmfright{o1}
      \fmflabel{$\mu,a$}{i1}
      \fmflabel{$\nu,b$}{i2}
      \fmflabel{$\rho,c$}{o1}
      \fmf{dbl_curly,label=$k$,label.dist=0.3cm}{i1,v1}
      \fmf{dbl_curly,label=$p$,label.dist=0.3cm}{v1,i2}
      \fmf{gluon,label=$q$,side=right,label.dist=0.3cm}{v1,o1}
    \end{fmfgraph*}
  \end{fmffile}
}
=g_s f^{abc} [g^{\mu\nu}(k-p)^\rho+g^{\nu\rho}(p-q)^\mu+g^{\rho\mu}(q-k)^\nu].
\end{equation}
\\
\\
\begin{multline}
\parbox{4cm}{
  \begin{fmffile}{feyn3}
    \begin{fmfgraph*}(30,25)
      \fmfleft{i2,i1}
      \fmfright{o1,o2}
      \fmflabel{$\mu,a$}{i1}
      \fmflabel{$\nu,b$}{i2}
      \fmflabel{$\rho,c$}{o1}
      \fmflabel{$\sigma,d$}{o2}
      \fmf{dbl_curly}{i1,v1}
      \fmf{dbl_curly}{v1,i2}
      \fmf{gluon,side=right}{v1,o1}
      \fmf{gluon,side=right}{v1,o2}
    \end{fmfgraph*}
  \end{fmffile}
}
= -ig^2_s [ f^{abe} f^{cde} (g^{\mu\rho}g^{\nu\sigma} - g^{\mu\sigma}g^{\nu\rho})
+ f^{ace} f^{bde} (g^{\mu\nu}g^{\rho\sigma} - g^{\mu\sigma}g^{\nu\rho})
\\
+ f^{ade} f^{bce} (g^{\mu\nu}g^{\rho\sigma} - g^{\mu\rho}g^{\nu\sigma})]
\end{multline}
Coupling between $A_5$ and gluon:
\\
\begin{equation}
\parbox{4cm}{
  \begin{fmffile}{feyn4}
    \begin{fmfgraph*}(25,25)
      \fmfleft{i2,i1}
      \fmfright{o1}
      \fmflabel{$a$}{i1}
      \fmflabel{$b$}{i2}
      \fmflabel{$\mu,c$}{o1}
      \fmf{dashes,label=$p$}{i1,v1}
      \fmf{dashes,label=$q$}{v1,i2}
      \fmf{gluon}{v1,o1}
    \end{fmfgraph*}
  \end{fmffile}
}
= -g_s f^{abc}(p-q)^\mu
\end{equation}
\\
\\
\begin{equation}
\parbox{4cm}{
  \begin{fmffile}{feyn5}
    \begin{fmfgraph*}(30,25)
      \fmfleft{i2,i1}
      \fmfright{o1,o2}
      \fmflabel{$a$}{i1}
      \fmflabel{$b$}{i2}
      \fmflabel{$\mu,c$}{o1}
      \fmflabel{$\nu,d$}{o2}
      \fmf{dashes}{i1,v1}
      \fmf{dashes}{v1,i2}
      \fmf{gluon}{v1,o1}
      \fmf{gluon}{v1,o2}
    \end{fmfgraph*}
  \end{fmffile}
}
= ig^2_s g^{\mu\nu}(f^{ace}f^{bde} + f^{ade}f^{bce})
\end{equation}
\\
The coupling between gluon, KK gluon and $A_5$:
\\
\begin{equation}
\parbox{4cm}{
  \begin{fmffile}{feyn6}
    \begin{fmfgraph*}(25,25)
      \fmfleft{i2,i1}
      \fmfright{o1}
      \fmflabel{$a,\mu$}{i1}
      \fmflabel{$b,\nu$}{i2}
      \fmflabel{$c$}{o1}
      \fmf{dbl_curly}{i1,v1}
      \fmf{gluon}{v1,i2}
      \fmf{dashes}{v1,o1}
    \end{fmfgraph*}
  \end{fmffile}
}
= -im_{KK} g_s f^{abc} g^{\mu\nu}
\end{equation}
\\
Coupling {{between ghost of KK gluon and gluon, where }}the ghost of gluon is denoted as dotted line, and the ghost of KK gluon is denoted as circle line:
\\
\begin{equation}
\parbox{4cm}{
  \begin{fmffile}{feyn7}
    \begin{fmfgraph*}(25,25)
      \fmfleft{i2,i1}
      \fmfright{o1}
      \fmflabel{$a$}{i1}
      \fmflabel{$b$}{i2}
      \fmflabel{$c,\mu$}{o1}
      \fmf{dbl_dots_arrow,label=$p$}{v1,i1}
      \fmf{dbl_dots_arrow}{i2,v1}
      \fmf{gluon}{v1,o1}
    \end{fmfgraph*}
  \end{fmffile}
}
\end{equation}
\\
\\
{{Coupling between ghost of KK gluon, ghost of gluon, and KK gluon:}}
\begin{equation}
\parbox{4cm}{
  \begin{fmffile}{feyn8}
    \begin{fmfgraph*}(25,25)
      \fmfleft{i2,i1}
      \fmfright{o1}
      \fmflabel{$a$}{i1}
      \fmflabel{$b$}{i2}
      \fmflabel{$c,\mu$}{o1}
      \fmf{dbl_dots_arrow,label=$p$}{v1,i1}
      \fmf{dots_arrow}{i2,v1}
      \fmf{dbl_curly}{v1,o1}
    \end{fmfgraph*}
  \end{fmffile}
}
= -g_s f^{abc}p^\mu
\end{equation}
\\
\\
\begin{equation}
\parbox{4cm}{
  \begin{fmffile}{feyn9}
    \begin{fmfgraph*}(25,25)
      \fmfleft{i2,i1}
      \fmfright{o1}
      \fmflabel{$a$}{i1}
      \fmflabel{$b$}{i2}
      \fmflabel{$c,\mu$}{o1}
      \fmf{dots_arrow,label=$p$}{v1,i1}
      \fmf{dbl_dots_arrow}{i2,v1}
      \fmf{dbl_curly}{v1,o1}
    \end{fmfgraph*}
  \end{fmffile}
}
= -g_s f^{abc}p^\mu
\end{equation}
\\

\section{Benchmark numbers for the virtual corrections}
\label{sec:bench}
To facilitate a convenient comparison to our calculation, we provide the explicit numbers of squared LO and NLO virtual amplitudes in this appendix for a single phase space point:
\begin{eqnarray}
  p_1 &=& ( 736.270321435165, 0,  0,   736.270321435165),
\nn
\\
p_2 &=& (736.270321435165, 0, 0,  -736.270321435165),
\nn
\\
p_3 &=& (736.270321435165, 0,  630.456435312587, -338.913586920306),
\nn
\\
p_4 &=& (736.270321435165, 0,  -630.456435312587,  338.913586920306),
\end{eqnarray}
where the momentum are given in the unit of GeV. The {{pole}} mass of top quark is chosen as $172.5$ GeV. The renormalization scale is set to be $\mu = 172.5$ GeV. Strong coupling constant is chosen as $\alpha_s(\mu)=0.107663194383306${{, corresponds to $\alpha_s(M_Z)=0.118$}}. The {{complex}} mass of KK gluon is {{given by}} $m^2_{KK} = 2250000 - 362039.269141765i \,(\mathrm{GeV}^2)$. Finally the coupling between the quark and KK gluon at the scale $m_t$ are
\begin{gather}
  C^q_L(m_t) = -1.74, \qquad C^q_R(m_t) = 1.74, \nn\\
  C^t_L(m_t) = 1.74, \qquad C^t_R(m_t) = -1.74.
\end{gather}
We defined the LO squared amplitude as
\begin{gather}
  \sum_{\rm spin,color}|\mathcal{M}^{(0)}_{\rm SM}|^2 = a,
\nn
\\
  \sum_{\rm spin,color}|\mathcal{M}^{(0)}_{\rm KK}|^2 = b,
\end{gather}
where $\mathcal{M}^{(0)}_{\rm SM}$ and $\mathcal{M}^{(0)}_{\rm KK}$ are the LO amplitude for $q\bar{q}\rightarrow t\bar{t}$ induced by gluon and KK gluon, respectively. Although the SM amplitudes are well known, we nonetheless give them here for the convenience of the reader. At the specific phase space point we have chosen, the numbers on the amplitudes are given in Table.~\ref{tab:lo}.

Next we define the virtual squared amplitudes as
\begin{gather}
  \sum_{\rm spin,color}2\mathrm{Re}(\mathcal{M}^{(0)}_{\rm SM}
\mathcal{M}^{(1)*}_{\rm SM}) = \frac{(4\pi)^\ep}{\Gamma(1-\ep)}
 \left( \frac{c_{-2}}{\ep^2_{\rm IR}} + \frac{c_{-1}}{\ep_{\rm IR}} + c_0
\right),
\nn
\\
  \sum_{\rm spin,color}2\mathrm{Re}(\mathcal{M}^{(0)}_{\rm KK}
\mathcal{M}^{(1)*}_{\rm KK}) = \frac{(4\pi)^\ep}{\Gamma(1-\ep)}
 \left( \frac{d_{-2}}{\ep^2_{\rm IR}} + \frac{d_{-1}}{\ep_{\rm IR}} + d_0
\right),
\end{gather}
where $\mathcal{M}^{(1)}_{\rm SM}$ is the SM one-loop virtual corrections, and $\mathcal{M}^{(1)}_{\rm KK}$ is the one-loop virtual corrections induced by KK gluon, {{including those diagrams in figures 3, 7, 8, and 9}}. All the scalar functions with complex argument are evaluated by the Fortran package {\sc OneLOop}~\cite{vanHameren:2010cp}. The divergent coefficients, $c_{-2}$, $c_{-1}$, $d_{-2}$ and $d_{-1}$ are given explicitly in a simple form in the text, so we will not {{present their numerical values here.}} The finite terms, $c_0$ and $d_0$ are given in Table.~\ref{tab:nlo}.
\begin{table}[h]
  \centering
  \begin{tabular}{|c|c|}
\hline
   \, $a\, ({\rm GeV}^{-2})$ \, & \, $18.55003354597$ \,
\\
\hline
    $b\, ({\rm GeV}^{-2})$ & $2896.300721184$
\\
\hline
  \end{tabular}
  \caption{Numbers on the LO squared amplitude at a specific phase space point.}
\label{tab:lo}
\end{table}
\begin{table}[h]
  \centering
  \begin{tabular}{|c|c|}
\hline
  \,  $c_0\, ({\rm GeV}^{-2})$ \,& \,$-8.084513116805$\,
\\
\hline
    $d_0\, ({\rm GeV}^{-2})$ & $-1966.611002687$
\\
\hline
  \end{tabular}
  \caption{Numbers on the NLO squared amplitude at a specific phase space point.}
\label{tab:nlo}
\end{table}
\bibliography{kk}{}
\end{document}